\documentclass[3p]{elsarticle}

\def\fighome {figures}
\def\fignewhome {figures}
\usepackage{pifont, natbib, geometry, fleqn}
\usepackage[flushmargin]{footmisc}
\usepackage{amsmath}
\usepackage{epsfig,epsf,psfrag,amssymb,amsfonts,latexsym,slashbox,graphicx,bm, xcolor,url}
\usepackage[caption=false,font=footnotesize]{subfig}%
\usepackage{cases}
\usepackage{verbatim}
\usepackage[mathscr]{eucal}
\usepackage{fixltx2e}
\usepackage{array}

\DeclareMathOperator{\supp}{supp}

\DeclareMathOperator{\unif}{Unif}

\def\lcx{{\,\,\underset{cx}{\leq}\,\,}}

\def\leqst{{\,\,\overset{st}{\leq}\,\,}}
\def\geqst{{\,\,\overset{st}{\geq}\,\,}}

\def\tha{{\mbox{\tiny th}}}
\DeclareMathOperator{\Var}{Var}

 \def\0{{\bf 0}}

\def\viz{{viz.,\ \/}}
\def\ie{{i.e.,\ \/}}
\def\eg{{e.g.,\ \/}}

\def\st{{s.t.  }}

\def\nn{\nonumber}

\def\qed{\hfill\hbox{${\vcenter{\vbox{
    \hrule height 0.4pt\hbox{\vrule width 0.4pt height 6pt
    \kern5pt\vrule width 0.4pt}\hrule height 0.4pt}}}$}}

\definecolor{myred}{rgb}{0.3,0.0,0.7}
\definecolor{dkg}{rgb}{0.1,0.7,0.2}
\definecolor{dkb}{rgb}{0.0,0.2,0.8}

\def\bfa{{\mathbf a}}

\def\bfx{{\mathbf x}}           

\def\bfz{{\mathbf z}} 
\def\bfA{{\mathbf A}}

\def\bfD{{\mathbf D}}

\def\bfI{{\mathbf I}}
\def\bfJ{{\mathbf J}}

\def\bfT{{\mathbf T}}

\def\bfV{{\mathbf V}}

\def\bfX{{\mathbf X}}
\def\bfY{{\mathbf Y}}
\def\bfZ{{\mathbf Z}}    

\def\pibf{\hbox{\boldmath$\pi$\unboldmath}}

\def\Pibf{{\bf \Pi}}

\def\Ac{{\cal A}}

\def\Lc{{\cal L}}

\def\Ebb{{\mathbb E}}

\def\Nbb{{\mathbb N}}

\def\Pbb{{\mathbb P}}

\newcommand{\bprf}{\begin{myproof}}
\newcommand{\eprf}{\end{myproof}}
\newcommand{\bp}{\begin{psfrags}}
\newcommand{\ep}{\end{psfrags}}
\newcommand{\bl}{\begin{lemma}}
\newcommand{\el}{\end{lemma}}
\newcommand{\bt}{\begin{theorem}}
\newcommand{\et}{\end{theorem}}
\newcommand{\bc}{\begin{center}}
\newcommand{\ec}{\end{center}}
\newcommand{\bi}{\begin{itemize}}
\newcommand{\ei}{\end{itemize}}
\newcommand{\ben}{\begin{enumerate}}
\newcommand{\een}{\end{enumerate}}
\newcommand{\bd}{\begin{definition}}
\newcommand{\ed}{\end{definition}}
\def\beq{\begin{equation}}
\def\eeq{\end{equation}\noindent}
\def\beqn{\begin{eqnarray}}
\def\eeqn{\end{eqnarray} \noindent}
\def\beqnn{  \begin{eqnarray*}}
\def\eeqnn{\end{eqnarray*}  \noindent}
\def\bcase{  \begin{numcases}}
\def\ecase{\end{numcases}   \noindent}
\def\bsbcase{  \begin{subnumcases}}
\def\esbcase{\end{subnumcases}   \noindent}

\def\defeq{{:=}}

\newtheorem{theorem}{Theorem}
\newtheorem{corollary}{Corollary}
\newtheorem{lemma}{Lemma}
\newtheorem{definition}{Definition}

\newtheorem{proposition}{Proposition}

\newenvironment{myproof}{\noindent{\em Proof:} \hspace*{1em}}{
    \hspace*{\fill} $\Box$ }
\newenvironment{proof_of}[1]{\noindent {\em Proof of #1:}
    \hspace*{1em} }{\hspace*{\fill} $\Box$ }

\def\psfancypar#1#2{\begingroup\def\par{\endgraf\endgroup\lineskiplimit=0pt}
               \setbox2=\hbox{\large\sc #2}
               \newdimen\tmpht \tmpht \ht2 \advance\tmpht by \baselineskip
               \font\hhuge=Times-Bold at \tmpht
               \setbox1=\hbox{{\hhuge #1}}
               \count7=\tmpht \count8=\ht1
               \divide\count8 by 1000 \divide\count7 by \count8 
               \tmpht=.001\tmpht\multiply\tmpht by \count7 
               \font\hhuge=Times-Bold at \tmpht
               \setbox1=\hbox{{\hhuge #1}}
               \noindent
                \hangindent1.05\wd1
               \hangafter=-2 {\hskip-\hangindent
               \lower1\ht1\hbox{\raise1.0\ht2\copy1}%
                \kern-0\wd1}\copy2\lineskiplimit=-1000pt}

\def\Kout{\setbox1=\hbox{\Huge\bf K}\hbox to
1.05\wd1{\hspace{.05\wd1}
}}
\def\Sout{\setbox1=\hbox{\Huge\bf S}\hbox to 1.05\wd1{\hspace{.05\wd1}
}}

\def\defeq{:=}

\def\perm{{\mbox{perm}}}

\def\fifo{{\mbox{\tiny FIFO}}}

\def\lbd2{\mbox{LB}_2}
\def\ubd2{\mbox{UB}_2}

\def\slbd2{\mbox{\tiny LB}_2}
\def\subd2{\mbox{\tiny UB}}

\def\fifo{{\mbox{\tiny FIFO}}}
\def\rand{{\mbox{\tiny RAND}}}

\def\spread{{V}}

\def\bfspread{{\bfV}}
 \def\policy{{\gamma}}
\def\Lc{{\mathcal{Q}}}
\def\psq{{\mbox{\tiny PS}}}
 \def\inft{{\mbox{\tiny INF}}}

\begin{document}

\begin{frontmatter}

\title{Seeing Through Black Boxes : Tracking Transactions through Queues under Monitoring Resource Constraints\tnoteref{t1}}

\tnotetext[t1]{The work is presented in part in
\cite{AnandkumarEtal:09MAMA}.}

\author[aa]{Animashree Anandkumar\corref{cor1}}
\ead{a.anandkumar@uci.edu}

\author[cb]{Ting He}
\ead{the@us.ibm.com}

\author[cb]{Chatschik Bisdikian}
\ead{bisdik@us.ibm.com}

\author[cb]{Dakshi Agrawal}
\ead{agrawal@us.ibm.com}

\address[aa]{EECS Dept., University of California, Irvine, CA 92697, USA.}

\address[cb]{Networking group, IBM Watson Research,
Hawthorne, NY 10532, USA.}

\cortext[cor1]{Corresponding author}

\begin{abstract}{\small The problem of optimal allocation of monitoring resources for
 tracking  transactions progressing through a distributed system, modeled as a  queueing network,  is considered. Two forms of monitoring information are considered, \viz
locally unique transaction identifiers, and arrival and departure
timestamps of transactions at each processing queue. The timestamps
are assumed available at all the queues but in the absence of
identifiers, only enable imprecise tracking  since parallel
processing can result in out-of-order departures. On the other hand,
identifiers enable precise tracking but are not available without
proper instrumentation. Given an   instrumentation budget, only   a
subset of  queues can be selected for production of identifiers,
while the remaining queues have to resort to imprecise tracking
using timestamps. The goal is then to optimally allocate the
instrumentation budget to maximize the overall tracking accuracy.
The challenge is that the optimal allocation strategy depends on
accuracies  of timestamp-based tracking at different queues, which
has complex dependencies on the arrival and  service processes, and
the queueing discipline.  We propose two  simple heuristics for
allocation by predicting the order of  timestamp-based  tracking
accuracies  of different queues. We derive sufficient conditions for
these heuristics to achieve optimality through the notion of
stochastic comparison of queues. Simulations show that our
heuristics are close to optimality, even when the parameters deviate
from these conditions.
 }\end{abstract}


\end{frontmatter}


\section{Introduction}

Transaction processing has been at the heart of information
technology since the 1950s when the first large online reservation
system went into operation \cite{Bernstein:book,Borr1981}. Today
transaction processing is at the core of enterprise IT systems
operated by telecommunication service providers, financial
institutions and virtual retailers. The scope of transaction
processing has widened to incorporate multiple software components
and applications, servers, middleware, backend databases, and
multiple information sources \cite{Spector&etal:2004IBM}.

The growing complexities  of transaction processing presents new
challenges to system management and support.  Today's support
helpdesks are no longer  knowledgeable with the intimate details of
transaction processing. The presence of heterogeneous components,
legacy systems and third-party ``black box" components
\cite{Aguilera2003} makes debugging, a slow and an expensive ordeal.
It is thus highly desirable to speed up debugging through automated
monitoring solutions.

Although tools   may be available   for independent trouble-shooting
within each of the components, they cannot capture the entire
life-cycle of a transaction, and thus cannot support diagnosis at
the transaction level. Instead,  an integrated end-to-end solution
which tracks the entire path of transaction processing is required
\cite{Thereska&etal:Sigmetrics06}. An end-to-end monitor collects
transaction records from different components and then correlates or
matches them to obtain the complete transaction path. If all the
components are instrumented properly, e.g., using techniques in
\cite{arm,Barham2004,Thereska&etal:Sigmetrics06}, then each
transaction record at every component is tagged with a unique
\emph{identifier} corresponding to the transaction generating it.
Using these identifiers, correlation of transaction records at
different components can then be done precisely.

In many practical scenarios, however, complete instrumentation of
all the components is rarely the norm. This is due to the presence
of legacy systems and third-party components with monitors producing
incompatible transaction records, which in effect, is a set of
``black boxes".  In the extreme case when none of the components is
instrumented, monitoring solutions have to  fall back on other
generic features in the records such as timestamps to statistically
``guess" the  set of records likely generated by the same
transaction, and thereby infer the path taken by that transaction
\cite{Aguilera2003,Anandkumar&Bisdikian&Agrawal:08Sigmetrics}, with
the caveat that the results may be erroneous.

Most real systems lie somewhere in the middle of the spectrum
between the extreme scenarios of fully instrumented and fully
non-instrumented systems. In fact, most system integration and
instrumentation is a gradual process which starts from an ensemble
of black boxes and slowly transitions to a system of   ``clear" or
``open" boxes as the support staff acquaint themselves with various
components. Given sufficient time and efforts, skilled programmers
are able to \emph{retrofit} instrumentation\footnote{\scriptsize
Note that with partial instrumentation here the identifiers are
local,  defined only within each queue, which is different from the
global identifiers in fully instrumented systems
\cite{Barham2004,Thereska&etal:Sigmetrics06}.} to components by
injecting monitoring code or building an extra layer of middleware
\cite{Thereska&etal:Sigmetrics06}. A complete instrumentation,
however, can incur daunting costs and is nevertheless wasteful in
components where statistical tracking using timestamps already has
good accuracy. Our goal is then to systematically characterize the
performance of partially instrumented monitoring systems and
identify components where retrofitting instrumentation is most
required.

We answer the following questions: given a limited budget for
instrumentation, what is the optimal allocation strategy to maximize
overall  accuracy of tracking transactions? What is the influence of
various system parameters, such as the queueing arrival and the
service rates, on the instrumentation strategy and  the  tracking
accuracy? Are there simple easy-to-implement heuristics that also
have good performance guarantees? What follows is a set of
systematic answers to these questions.

\begin{figure}[tb]
\bc\bp\psfrag{Q0}[l]{\scriptsize $Q_0$}\psfrag{Q1}[l]{\scriptsize
$Q_1$}\psfrag{Q2}[l]{\scriptsize $Q_2$}\psfrag{?}[c]{\scriptsize ?}
\psfrag{S4}{ }
\includegraphics[width=2.7in]{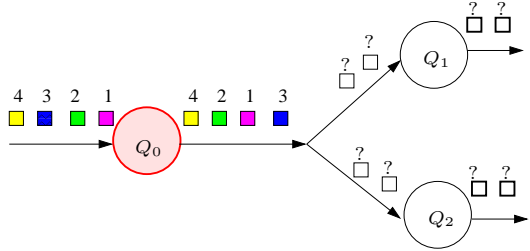}\ep\caption{\scriptsize
Introducing identifiers to timestamps at queue $Q_0$ through
instrumentation  precisely tracks transactions progressing through
it. On the other hand, non-instrumented queues $Q_1$ and $Q_2$  have
to track transactions using only  arrival and departure timestamps
  may incur errors due to uncertainty in the order of departures.}\label{fig:problemsetup} \ec\end{figure}

\subsection{Technical Approach and Contributions}

We consider the problem of tracking transactions through a
distributed system with limited instrumentation support. Our goal is
to select an optimal subset of components for instrumentation under
a budget constraint such that when combined with statistical
tracking (using timestamps) at the non-instrumented components, the
overall tracking accuracy is maximized.

Our contributions are three fold. First, we analyze the accuracy of
statistical tracking using timestamps at a queue and characterize
its dependency on different queueing parameters. Second, using these
insights, we propose two simple heuristics for the  instrumentation
allocation problem. Third, we   derive sufficient conditions for
these heuristics to achieve optimality, based on the arrival and the
service distributions at the queues.

{\bf Model: }We model the progress of the transactions in a
distributed system as a  queueing network, where each queue
represents a  system component. By default, we assume the
availability of (an ordered) set of arrival and departure timestamps
at each queue while identifiers are only available upon
instrumentation (queue  $Q_0$  in  Fig.\ref{fig:problemsetup}).  Due
to parallel processing of transactions, \eg in  infinite server or
processor-sharing queues\footnote{\scriptsize There  is no
uncertainty in the order of departures  for single-server queues
with fixed order processing. Hence, their timestamp-based tracking
is error-free, and we do not  consider them    for allocation.}, the
order of departures is not unique, and in the absence of
identifiers, tracking transactions through a queue requires
statistical matching techniques.  We analyze tracking accuracies
using timestamps under two simple statistical matching policies.
Identifiers are available only upon instrumentation and by
instrumenting a queue, we mean  injecting code  or building a
middleware wrapper   which  tags each timestamp with an identifier
unique to the transaction,  leading to error-free tracking at those
queues.

{\bf Formulation:} Based on the above model, we formulate  a
resource allocation problem, where we optimally allocate the
available amount of monitoring   resources by selecting queues for
instrumentation such that   the overall tracking accuracy   is
maximized.   The   optimal   allocation strategy thus    selects
queues for instrumentation in the increasing order of their
timestamp-based tracking accuracies, until the budget constraints
are met.    However, the exact expression  of tracking accuracy at
each non-instrumented queue is not tractable to compute in general,
and has complex dependencies on the arrival and service statistics,
and also on the queueing discipline.

{\bf Heuristic Solutions: } To overcome this obstacle, we propose
two simple heuristics for instrumentation allocation which predict
the order of the timestamp-based tracking accuracies at different
queues without computing the exact expressions. The first heuristic
predicts that the order of tracking accuracies   is in the reverse
order of their queueing load factors. The second heuristic predicts
the order of accuracies using an approximation for the tracking
accuracy, which becomes tight in the light load regime. The two
heuristics represent different tradeoffs in that  the load-factor
heuristic requires only the knowledge of the queueing load factors
while the approximation-based heuristic requires the full knowledge
of arrival and service processes but   is a more efficient
allocation strategy (demonstrated through both  theory and
simulations).

{\bf Optimality conditions: }We   provide sufficient conditions for
these heuristics to achieve optimality, i.e., to correctly rank  the
order of the tracking accuracies, based on the notions of
\emph{stochastic} and \emph{convex orders} of the arrival and
service   distributions of the queues. The conditions have intuitive
explanations in terms of the rate and the ``variability" of arrivals
and services. In particular, these heuristics are always optimal
when  all the arrival distributions and all the service
distributions  belong to the same family.  Simulations verify the
optimality of our heuristics under the derived conditions and also
show that our heuristics are close to optimality even when the
parameters deviate from these conditions.


 {\bf Alternative Formulation: }Besides allocating instrumentation resources, our heuristics  are
also applicable  in other scenarios of monitoring. For instance, for
a large system, the overhead in collecting  timestamp records  from
all the components may be too large. In this case, the optimal
monitoring resource allocation is to \emph{a priori}  select only a
subset of components (queues) with the highest timestamp-based
tracking accuracies  for data collection.  Our heuristics and their
optimality guarantees are directly applicable here.

{\bf Non-goals: }We   emphasize some of our ``non-goals". Our
formulation and solutions have a strong theoretical foundation and
are meant to provide guidelines for efficient instrumentation or
 data collection in different scenarios. We do not attempt
to replace existing   instrumentation-based monitoring tools (See
Section~\ref{sec:related} for a discussion) and exploit them when
available. Our belief is that existing monitoring solutions will
have broader application by allowing for partial instrumentation,
and we have a systematic approach for pursuing it.  Moreover,  our
solutions are not meant to automatically diagnose or correct faults,
characterize overall system performance, or provide real-time
analysis, although such exercises can be carried out after
monitoring the transaction paths.



\subsection{Related Work}\label{sec:related}

The early literature on monitoring distributed systems relies on
deep understanding of internal system structures so that
instrumentation code can be injected into proper places to record
system activities at process or object levels
\cite{JoyceEtal:87TOCS,Chandy1985,Mansourisamani1993}. These
solutions become difficult to implement in modern systems where
components are typically developed independently. Most  existing
monitoring solutions rely on certain types of instrumentation that
can expose the activities of interest
\cite{Schmid2007,Chen2002,Chen2004,Thereska&etal:Sigmetrics06}.
There are also a number of commercially-available products for
monitoring and trouble shooting in distributed systems
\cite{arm,Aguilera2003,Liu2007}, which are again based on
instrumenting the system software.

While instrumentation provides reliable monitoring information, it
has limited use in heterogeneous systems where many components are
from third-party vendors or legacy systems. One approach is to make
the instrumentation as component-independent as possible, e.g., by
limiting changes to system code rather than user-space code
\cite{Tak&etal:09USENIX}. Another approach is to treat     each
component   as a black box and only rely on external activities of
these black boxes for monitoring
\cite{Aguilera2003,Liu2007,Barham2004}. These existing black-box
based solutions can be divided into two approaches: identifier-based
approach \cite{Barham2004} which tags each incoming transaction with
a unique identifier that is associated with it throughout the
system, converting the problem to the instrumented case, and
trace-based approach \cite{Aguilera2003,Liu2007} which uses
statistical techniques to extract monitoring information from
non-tagged activities. For example, \cite{Aguilera2003,Liu2007} use
messages between components to infer causal paths and bottlenecks.
We share a similar view as \cite{Aguilera2003,Liu2007} in  that a
monitoring solution should be as non-intrusive and agnostic as
possible to allow for broad application, especially in systems
involving black boxes, but there are two key differences that
distinguish our work from this literature: (i) we are interested in
monitoring individual transactions rather than aggregate system
behaviors such as causal paths and bottlenecks, and (ii) we take a
hybrid approach of using both passive monitoring (via
timestamp-based tracking) and instrumentation (that introduces
identifiers), but treat the latter as a limited resource   to be
allocated judiciously.

In  \cite{Anandkumar&Bisdikian&Agrawal:08Sigmetrics},   tracking of
individual transactions in a distributed system based solely on
timestamps is considered. However,
\cite{Anandkumar&Bisdikian&Agrawal:08Sigmetrics}  focuses on
developing optimal matching policies for timestamp-based transaction
monitoring, whereas we focus on   the comparison of tracking
accuracies at different subsystems while leveraging statistical
matching policies discussed in
\cite{Anandkumar&Bisdikian&Agrawal:08Sigmetrics}  for tracking in
the non-instrumented states. The stochastic comparison techniques
used in this paper has a rich history and has been applied compare
different queueing parameters such as delay and throughput \cite[Ch.
14]{Shaked&Shanthikumar:book94}.   To the best of our knowledge,
comparison of monitoring accuracies at different queues has not been
considered before.


\emph{Organization: }The paper is organized as follows. In Section
\ref{sec:model}, we describe the system model and problem
formulation.  In Section \ref{sec:monitoring}, we analyze the
policies for matching timestamps. In Section \ref{sec:2approaches},
we propose the two heuristics for monitoring resource allocation.
In Section \ref{sec:stoccomp}, we introduce the notion of stochastic
comparison. In Section \ref{sec:comp}, we  derive   sufficient
conditions for the  optimality of the two heuristics for network of
infinite-server queues.  Section \ref{sec:bcmp} deals with
extensions to  general product-form queues.  In Section
\ref{sec:sim}, we  evaluate the efficiency of heuristics through
simulations.  Section \ref{sec:conc} concludes our paper.

\section{System Model and Formulation}\label{sec:model}

We now describe the queueing model in detail and then formulate the
problem of optimal monitoring resource allocation. Before we
proceed, here are a few comments regarding the notation used in this
paper. Vectors are represented by boldface, \eg $\bfX$ and $X(i)$ is
its $i^{\tha}$ element. Let $f_X(x)$, $F_X(x)$ and $\bar{F}_X(x)$
denote the probability density function (pdf), cumulative
distribution function (cdf) and complementary cumulative
distribution function (ccdf) of a continuous variable $X$. Let
$\Ebb[X]$ denote its expectation and let $\supp(f_X)$ denote the
support of $f_X$.

\subsection{System Model}

We consider a queueing network, and initially limit to the  case
where all the queues are infinite server $(GI/GI/\infty)$. The
arrival and service times are drawn i.i.d. from  general continuous
pdfs $f_{X}$ and $f_{T}$. In Section \ref{sec:bcmp}, we  generalize
some of our results to the product-form queues. We assume that the
sequence of queues visited  by each transaction is a Markov chain,
and the service is independent of the transition sequence.  The list
of notations for different queueing parameters is given in Table
\ref{table:listofsymbols}. The propagation delays and
synchronization errors between different queues are assumed
 independent of the service or arrival realizations.

Given a set of ordered arrival and departure timestamps, $\bfY_k$
and $\bfD_k$ at queue $Q_k$, there is a relationship between the
service times and the true matching $\pibf^t_k$ between the arrivals
and the departures, as \beq  T_k(i)  =  D_k(\pi_k^t  (i)) -
Y_k(i), \quad i\in \Nbb.\label{eqn:T}\eeq Hence,   $\pi_k^t(i)$ is
the rank of a departure timestamp corresponding to the $i^{\tha}$
arrival to the queue $Q_k$. Since we have access to only the arrival
and departure timestamps $\bfY_k$ and $\bfD_k$, and \emph{not} to
the actual service times $\bfT_k$, the true matching $\pibf_k^t$ is
unknown. A \emph{bipartite matching policy} $\policy$ comes up with
a probable  matching $\pibf^\policy$ between the arrival timestamps
$\bfY_k$ and the departure timestamps $\bfD_k$, which  yields
correct matchings with a certain degree of accuracy, and is
discussed in detail   in Section \ref{sec:monitoring}. In addition,
we assume that identical policies $\gamma$ are employed for matching
at all the queues to facilitate comparison of their tracking
accuracies.

\begin{figure}[t]\hfil\subfloat[a][Busy Period $B=1$]{
\begin{minipage}{1.5in}\bc\bp\psfrag{D(1)}[c]{\scriptsize $D(1)$}
\psfrag{Y(1)}[l]{\scriptsize$Y(1)$}
\psfrag{Y(2)}[l]{\scriptsize$Y(2)$}
\psfrag{X(1)}[c]{\scriptsize$X(1)$}\psfrag{T(1)}[c]{\scriptsize$T(1)$}
\psfrag{Time}[l]{\small Time}\fbox{
\includegraphics[width=1.3in,height=1in]{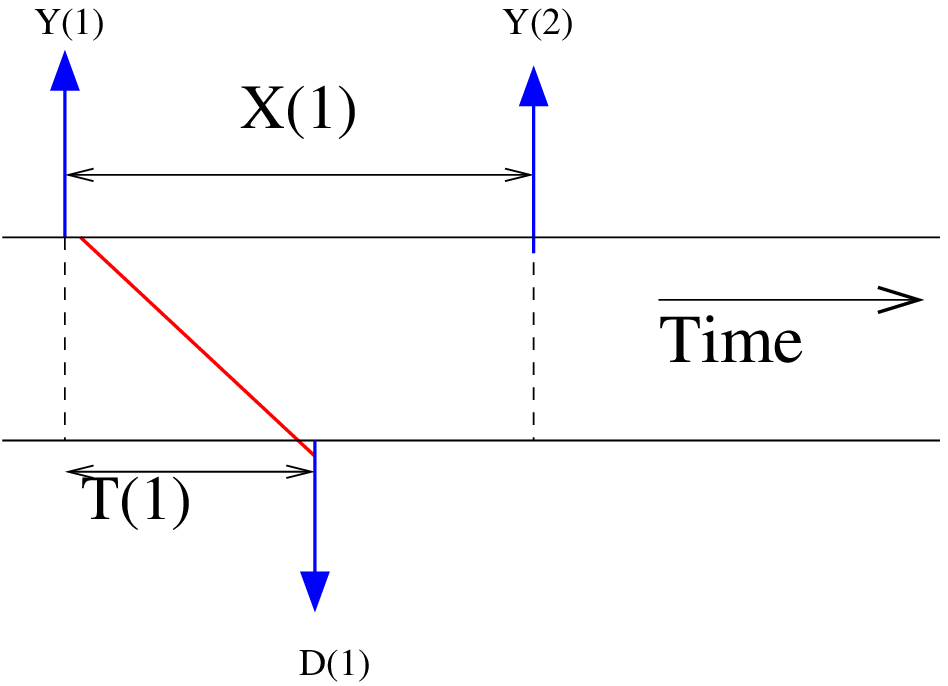}}\ep\ec\end{minipage}}
\hfil\subfloat[b][Busy Period $B=2$]{
\begin{minipage}{1.5in}\bc\bp
\psfrag{Y(1)}[l]{\scriptsize$Y(1)$}\psfrag{Y(2)}[l]{\scriptsize$Y(2)$}
\psfrag{Y(3)}[l]{\scriptsize$Y(3)$}
\psfrag{X(1)}[c]{\scriptsize$X(1)$}\psfrag{X(2)}[l]{\scriptsize$X(2)$}
\psfrag{Time}[l]{\small Time}\psfrag{T(1)}[l]{\scriptsize$T(1)$}
\psfrag{T(2)}[l]{\scriptsize$T(2)$}\psfrag{Y0(2)}[l]{}
\psfrag{Y0(3)}[l]{} \psfrag{D(1)}[c]{\scriptsize $D(1)$}
\psfrag{D(2)}[l]{\scriptsize $D(2)$}
\fbox{\includegraphics[width=1.5in,height=1in]{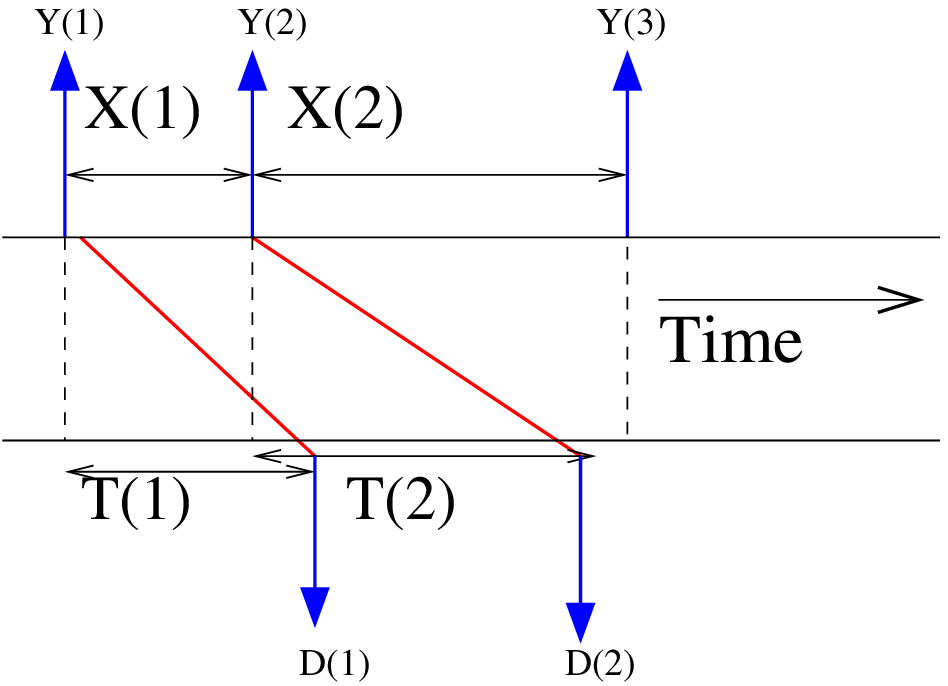}}\ep\ec\end{minipage}} \hfil
\caption{Random arrivals and departures lead to random busy period
sizes.}\label{fig:batchsizes}\end{figure}

Our analysis will be on a typical busy period, \ie a period of time,
starting from an empty queue until the next time  the queue becomes
free, as shown in Fig.\ref{fig:batchsizes}. Let $P^\policy(k)$ be
the  probability that the policy outputs a correct matchings between
{\em all} the arrivals and departures in a typical busy period at
queue $Q_k$.  We use $P^\policy(k)$ as the measure of
timestamp-based tracking accuracy,  given by \beq P^\policy(k) =
\sum_{b=1}^\infty \Pbb[ \pibf^\policy = \pibf^t,
B_k=b].\label{eqn:acc}\eeq


\begin{table}[t]
\begin{tabular}{|l|l|}
\hline $\bfX_k$ & vector of i.i.d inter-arrival times  \\ $\bfT_k$ &
vector of i.i.d service times
 \\   $\lambda_k
\defeq \frac{1}{\Ebb[X_k(1)]}$ & arrival rate  \\ $\mu_k  \defeq \frac{1}{\Ebb[T_k(1)]}$  &service
rate \\ $\rho_k \defeq \frac{\lambda_k}{\mu_k}$& load factor \\
$\spread_k$& $T_k(1) - T_k(2):$ spread of service time\\ $\bfY_k$&
vector of  arrival times
\\ $\bfD_k$&vector of  departure times \\ $\pibf^t_k$ & true
matching  btw.   arrivals \& departures
\\ $\pibf^\policy_k$ &   matching according to policy $\policy$\\ $B_k$ & (random)  no. of arrivals in a busy period
\\
$P^\gamma(k)$ & prob. of correct matching in a busy period\\
$P_b^\policy(k)$ & cond. prob. of correct match given
$B_k=b$\\\hline
\end{tabular}
\caption{Symbol list. Subscript $k$ means queue
$Q_k$.}\label{table:listofsymbols}\end{table}


\subsection{Problem Formulation}

We are now ready to state the problem of optimal monitoring resource
allocation. Given a budget constraint of instrumenting at most $E$
number of queues to enable    precise tracking through the
production of identifiers, our goal is to select $E$ number of
queues  in $\Lc$ such that the overall tracking accuracy is
maximized. For each queue $Q_k$, let $z_k\in \{0,1\}$ be the
indicator if it is selected for instrumentation. Then, the effective
tracking accuracy at queue $Q_k$ after instrumentation decisions is
\[ z_k + (1-z_k) P^\policy(k),\] since the tracking accuracy is
unity  when identifiers are available and $ P^\policy(k)$ is the
accuracy based on  using only  timestamps under a statistical
matching policy $\policy$. Formally, the optimization is
\begin{align}    \bfz_*(E;\Lc )&\defeq \arg\max_{\bfz} \sum_{Q_k \in
\Lc }  \{z_k + (1-z_k)
 P^\policy(k)\}, \label{eqn:onestepopt}\\ \st\quad \sum_{Q_k \in
\Lc }z_k   &\leq E,\quad z_k\in \{0,1\},\bfz\defeq\{z_k: Q_k\in
\Lc\}.\nn\end{align}  We can see that the optimal allocation
strategy is to select $E$ number of queues with the lowest
timestamp-based tracking accuracies $P^\policy$.  The challenge, as
we will see, is in finding the tracking accuracy $P^\policy$ since
it has complex dependencies on the arrival and service processes.

\section{Timestamp-based Tracking}\label{sec:monitoring}

In this section, we describe the matching policies $\gamma$ employed
for associating the arrival and the departure timestamps at a queue,
and perform some preliminary analysis on the tracking accuracy of a
policy.

\subsection{Bipartite Matching Policies}\label{sec:match}

We now briefly describe two matching policies $\policy$ that can be
employed to match timestamps in the absence of identifiers, \viz the
first-in first-out (FIFO) rule and the random matching rule. The
relative performance of these policies depends on the arrival and
service statistics. These policies are \emph{non-parametric}, in the
sense that they require minimal knowledge about the  service
statistics for implementation.

Perhaps the simplest matching rule between the arrival  and
departure timestamps is the FIFO rule, which is an in-order matching
rule, \ie for a given busy-period size $B=b$, we have a fixed rule
$\pibf^\fifo= \bfI$, where $\bfI\defeq[1,2,\ldots]^T$ is the
identity vector.     The FIFO matching rule is fully
distribution-free: it does not require the knowledge of arrival or
service distribution and is always valid. By valid, we mean that the
FIFO match has a strict positive likelihood of being the true match
between the arrivals and the departures.     An expression for the
expected matching accuracy under   FIFO rule can be found in~\ref{proof:fifoexpect}.

In addition to  the FIFO matching rule, we  consider another simple
rule called random matching, where given a realization of arrivals
and departures in a busy period, we uniformly pick a valid  matching
among all possible matchings. The  random matching rule is almost
distribution-free: it only requires the knowledge of  $
\supp(f_{T_k})$, the support  of the  service pdf, in order to
ensure the validity of different matchings. This is because a valid
matching $\pibf$ at queue $Q_k$ in a busy period of size $B_k=b$
satisfies
 \beq \pibf:
\prod_{i=1}^b f_{T_k}[D_k(\pi(i )) -
Y_k(i)]>0,\label{eqn:validmatchset}\eeq  and the above expression
only requires the knowledge of the support bounds. An expression for
tracking accuracy $P^\rand$ under random matching is given in~\ref{app:random}.

 In contrast to the non-parametric FIFO and random matching rules,  the parametric {\em
maximum-likelihood} matching rule
\cite{Anandkumar&Bisdikian&Agrawal:08Sigmetrics} requires the full
knowledge of the service distribution. The maximum-likelihood rule
is defined as the rule  which maximizes the probability of correctly
matching all the arrivals and departures.  However, it is not
tractable to analyze this rule since it is fully adaptive to the
realization of arrivals and departures, and depends on the arrival
and service statistics in a complex manner. In many cases, the
simple  FIFO and random matching policies coincide with the
maximum-likelihood rule or have close to optimal performance, as
discussed below.

The effectiveness of using the FIFO or the random policy crucially
depends on the nature the service distribution (for a given
realization of arrivals).  For instance, under light-tailed
services, the probability of out-of-order departures is small and
hence, the FIFO rule is expected to have good tracking accuracy. In
fact, for Weibull\footnote{\scriptsize The pdf of a Weibull variable
is $f(x)=(\frac{w}{c}) (\frac{x}{c})^{w-1} \exp(-(\frac{x}{c})^w)$
for $x>0$, where $w$ and $c$ are   shape   and scale parameters.
When  $w>1$, the distribution
 is light tailed, when $w<1$, it is heavy tailed and $w=1$ is the exponential distribution.} family of distributions, with shape
parameter greater than one (and hence, light tailed), FIFO is the
optimal matching policy coinciding with    the maximum-likelihood
rule. More generally, FIFO rule is   optimal   whenever the service
pdf is log-concave \cite{Anandkumar&Bisdikian&Agrawal:08Sigmetrics}.

For heavy-tailed distributions, on the other hand, the chances of
out-of-order departures are high, and the    FIFO rule is not close
to the maximum-likelihood rule. In this case,   the random matching
rule may have better tracking accuracy  than the FIFO rule. This is
observed in our  simulations in
Fig.\ref{fig:singlelink_heavyWeibull} for Weibull distribution with
shape parameter smaller than one. Moreover, random matching is
optimal in case of batch arrivals to the infinite-server queue where
all possible matchings between the arrivals and departures are
equally likely, although we do not study this scenario in the paper.
Hence, the relative performance of FIFO and random matching rule
depends on the  service distribution.

\subsection{Tracking Accuracy}\label{sec:accuracy}

\begin{figure}[tb]\bc\bp\psfrag{Transition Link}[l]
{\small Transition Link $e$}\psfrag{Time}[l]{\small
Time}\psfrag{System}[l]{\small System}\psfrag{Occupancy}[l]{\small
Occupancy} \psfrag{S_0}[l]{\small $Q_0$}\psfrag{S_1}[l]{\small
$Q_1$}\psfrag{Busy Period}[l]{\small Busy
Periods}\psfrag{(Batch)}[l]{}\psfrag{Arrivals}[l]{\small Arrivals}
\psfrag{Departures}[l]{\small Departures}\psfrag{1}[l]{\scriptsize
$1$}\psfrag{2}[l]{\scriptsize $2$}\psfrag{0}[l]{\scriptsize $0$}
\includegraphics[width=2.7in,height=1.4in]{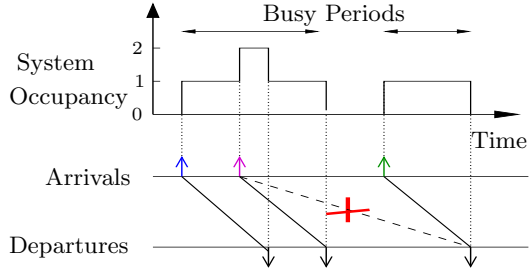}\ep\ec
\caption{Matching  arrival and departure timestamps  decomposes
across different  busy
periods.}\label{fig:2state_combined}\end{figure}

Recall we consider the probability of matching all timestamps in a
typical busy period to be the measure of tracking accuracy. Perhaps,
a more   straightforward measure of accuracy is  the probability of
correctly matching only a typical pair of arrival and departure
timestamps. This  however depends on    the probability of correctly
matching other arrivals and departures. On the other hand,  the
matching across busy periods  is independent, since a valid matching
between arrival and departure timestamps  occurs only within busy
periods not across them.  See Fig.\ref{fig:2state_combined}. Hence,
the probability of correct matching in a typical busy period
$P^\policy$ is the relevant measure for tracking accuracy.

The challenge is in computing $P^\policy$ in \eqref{eqn:acc}.
Consider FIFO matching as an example. Its accuracy is equal to (see~\ref{proof:fifoexpect})
\[P^{\fifo}\!\! \!\!\!= \sum_{b=1}^\infty\Pbb(\bigcap_{i =1}^{b-1}\{ T(i) \in [X(i), X(i) + T(i+1)]\} \cap
\{T(b) < X(b)\}),\] where the   events $T(i) \in [X(i), X(i) +
T(i+1)]$ and $T(b) < X(b)$ cannot be evaluated separately since are
correlated with one another other. We can see that the expression
becomes intractable as we increase $b$, the size of the busy period.

More generally, a matching policy $\policy$ may  select any one of
the valid matchings or permutations with a certain probability, and
the tracking accuracy $P^\gamma$ from \eqref{eqn:acc} becomes
\[P^\gamma = \sum_{b=1}^\infty \sum_{\pibf_j }\Pbb[\pibf^\gamma=\pibf^t=\pibf_j, B=b],\] where
the sum is over all the permutation vectors $\pibf_j$ over
$\{1,2,\ldots, b\}$. Since there are $b!$ number of permutation
vectors, we require exponential number of computations in $b$.

It is therefore not tractable to compute the tracking accuracies
$P^\policy(k)$ at different queues $Q_k$, in order to find  the
optimal resource allocation strategy in \eqref{eqn:onestepopt}.
Moreover, it is useful to obtain some general guidelines about the
influence of different queueing parameters on the resulting tracking
accuracy. Fortunately, we note that we do not need to know the exact
accuracies at different queues in the network to obtain the optimal
solution to instrumentation allocation in \eqref{eqn:onestepopt}. In
fact, it suffices to know the relative order of these accuracies.
The goal of this paper is to establish simple heuristics that can be
used to infer the order of matching accuracies without directly
computing them. To this end, we now propose two approaches with
different complexities and generality.  Later in
Section~\ref{sec:stoccomp} and \ref{sec:comp}, we derive sufficient
conditions for these heuristics to achieve optimality according to
\eqref{eqn:onestepopt}.

\section{Two Heuristics for Optimal Resource Allocation}\label{sec:2approaches}

 We propose two approaches to instrumentation allocation through
 prediction of   the order of timestamp-based
tracking accuracies $P^\policy(k)$ at different queues $Q_k$. One
approach is to  avoid computation of $P^\policy(k)$   altogether and
instead infer their order through simple queueing parameters such as
the load factors. The other approach is to approximately compute $
P^\policy(k)$ by only considering small busy-period sizes. Both
these simple approaches instrument queues independent of the
policies $\gamma$ employed for timestamp matching. We now describe
these two approaches in detail.

\subsection{Approach 1: Order of Load Factors}\label{sec:loadheuristic}

The load factor $\rho_k=\frac{\lambda_k}{\mu_k}$ of a queue  $Q_k$,
which is the ratio of the   arrival rate $\lambda_k$ to the
service rate $\mu_k$,   is perhaps the most commonly used queueing
parameter for performance evaluation of queues.  We propose the
load-factor heuristic for instrumentation allocation which selects
queues for instrumentation in the decreasing order of their load
factors  until the budget constraint is met. The load-factor
heuristic is   \emph{robust} since the selected set of  queues  is
invariant  under small perturbations in the arrival and service
statistics.

The load-factor heuristic   predicts queues with higher load
factors to have lower timestamp-based tracking accuracies.  This is
intuitive since a lighter load implies   a smaller number of
simultaneously-served  arrivals in the infinite-server queue on
average leading to a  lower uncertainty in the order of departures.
The intuition, however, does not extend when we consider queues with
different arrival and service distributions. The arrival and service
processes   influence the tracking accuracy in a complex manner, and
the load factor may not always capture the required effects for
comparison of tracking accuracies at different queues.

A simple example is two queues with same arrival rate,  one with
uniform service  $\unif(0,2m)$ on support $[0,2m]$ and the other
with deterministic service  of value $m_d > m$. Here, the
load-factor heuristic incorrectly predicts  the deterministic
service to have worse tracking accuracy, while, in fact, it
actually has perfect accuracy. Hence, the load-factor heuristic is
not universally optimal for instrumentation allocation.

An intuitive reason for the sub-optimality of the load-factor
heuristic  is that  there are two sources of errors impacting   the
tracking accuracy:   variability in service times leading to
uncertainty in the order of departures and high load factor
resulting in more simultaneous servicing in infinite-server queues
on average. The load-factor heuristic only captures the latter
effect and completely ignores the former. As we saw  in the above
example, simultaneous servicing does not always lead to bad accuracy
and is also governed by  the variability in the service times.

In many cases, different subsystems in a distributed system may have
similar   service distributions (such as from the same family), but
with different load factors. Here, the    load-factor heuristic may
correctly predict the order of the tracking accuracies.  We  prove a
sufficient set of conditions for   the optimality of the load-factor
heuristic in Section~\ref{sec:comp} by precisely  investigating the
dependency of the arrival and  the service processes on the tracking
accuracy.

\subsection{Approach 2: Small-Batch Approximation} \label{sec:eval}

The load-factor heuristic described in the previous section avoids
computation of the tracking accuracy altogether. We now propose an
alternative heuristic which approximates tracking accuracy through a
simple expression, and makes instrumentation decisions based on the
approximation. We later demonstrate the superiority of this
heuristic over the load-factor heuristic, both through theory and
simulations.

The approximation for tracking accuracy is based on the series
expansion \beq  P^\policy(k)  = \Pbb[B_k=1]+ \sum_{b=2}^\infty
P^\policy_b(k) \Pbb[ B_k=b ],  \label{eqn:series_onestep}\eeq where
$P_1^\policy=1$ since  when there is only  one transaction in the
busy period, tracking is perfect. Under sufficient variability of
the service times  (\ie not deterministic services), the probability
of correct matching typically decays with the busy-period size,
\[\lim_{b\to\infty} P^\policy_b(k)=0,\] since the number of possible matchings
grows exponentially with the busy-period size $b$ and  we make an
error almost surely  as the busy period size goes to infinity.
Hence, the   terms corresponding to larger busy-period sizes in
(\ref{eqn:series_onestep}) can be dropped and an approximate
tracking accuracy can be efficiently computed by limiting to small
busy-period sizes.


The simplest approximation is when we ignore all the terms in
(\ref{eqn:series_onestep}) except for the first one, which is simple
to evaluate. We refer to  this as the \emph{unit-batch}
approximation and use it to allocate instrumentation resources to
queues.  Note that the unit-batch approximation is slightly more
complex than the load-factor heuristic.  We   demonstrate, both
through theory and simulations,  that this leads to superior
performance  over the load-factor heuristic; the intuition being
that this heuristic captures additional features of the arrival and
service statistics.

At low arrival rate, this approximation (and also more refined ones
with more terms) becomes tight in the limit. Intuitively, at low
arrival rates, the dominant event is having a single arrival in each
busy period since the arrivals are widely separated on average.

\begin{proposition}\textsc{(Tightness at Low Arrival Rate).} As the arrival rate  to a queue
$Q_k$ goes to zero,   and the service distribution is kept fixed, we
have \beq \lim_{\lambda_k\to
0}\frac{\Pbb[B_k=1]}{P^\policy(k)}=1.\label{eqn:unitbatch}\eeq
\end{proposition}

\bprf As $\lambda_k\to 0$, we have $\Pbb[B_k=1] = \Pbb[X_k> T_k]\to
1$ and $P^\policy\to 1$ since the probability of out-of-order
departures goes to zero.
\eprf

%
%

Hence, the tracking accuracy $P^\gamma$ is well approximated by the
probability of unit busy period in the low arrival rate or the light
load regime. However, simulations in Section~\ref{sec:sim} show that
the unit-batch approximation correctly captures the trend of
$P^\gamma$ and is hence, an efficient strategy for instrumentation
allocation over a wider regime of loads.

\section{Preliminaries: Stochastic Comparison}\label{sec:stoccomp}

We have so far proposed two simple heuristics for optimal
instrumentation resource allocation which circumvent the challenges
in computing the tracking accuracies at various queues. Our goal is
to establish a general set of conditions on the arrival and service
processes, under which these simple heuristics coincide with the
optimal allocation strategy. To this end, we introduce the notion of
stochastic comparison of random variables.

Perhaps the simplest notion of comparing two random variables is
through their mean values. But very often, this comparison turns out
to be too loose to draw useful conclusions since the probability
distribution of the two variables can be very different. In the
context of this paper, comparing only queueing load factors, which
is just the average  system behavior, is not enough to always
guarantee an order of the tracking accuracies of the queues and
hence, optimality of the load-factor heuristic for instrumentation
allocation.

Instead, we impose stronger constraints on the distributions of the
variables under comparison to obtain useful conclusions.
 Here, we employ two  notions of stochastic comparison, \viz the stochastic
order and the convex order. The stochastic order is a stronger form
of comparing the mean values, while the convex order is a stronger
form of comparing the variances of random variables. The detailed definitions are given in~\ref{app:stocorder}.
 We use these
notions in Section~\ref{sec:comp}  to compare tracking accuracies at
different queues, and to derive sufficient conditions for the
optimality of the two proposed heuristics for resource allocation.

\subsection{Stochastic Comparison of Busy Periods}

We now provide some preliminary results on comparing the busy-period
sizes of queues under stochastic or convex orders of arrival and
service processes. We use these results in Section~\ref{sec:comp}
to obtain an order on the tracking accuracies of the queues thereby
establishing the optimality of our heuristics for instrumentation
allocation.


We now show that under a stochastic order of arrival processes and
service processes at two queues, we can guarantee a stochastic order
of the size of their  busy periods.

\bl\label{lemma:busyperiod}\textsc{(Comparison of Busy Periods under
Stochastic Order).} For two $GI/GI/\infty$ queues $Q_k,Q_m$ with
i.i.d arrivals $X_k, X_m$ and i.i.d  service times $T_k,T_{m}$, we
have \beq X_k \leqst X_m, T_k \geqst T_m \Rightarrow  B_k  \geqst
B_{m}.\label{eqn:lemmabatchstuncond}\eeq \el

\bprf  See~\ref{proof:busyperiod}  \eprf


The above result confirms our intuition that the   size of the busy
period    increases with faster arrivals and slower services (and
hence,  higher load factors), formalized   under the   notion of
stochastic order.

We now  consider an alternative scenario where   one queue has a
higher (normalized) service variability than the other,  formalized
by the presence of a convex order. We show that this also implies a
stochastic order on their busy-period sizes for the special case of
Poisson arrivals at all the queues.

\bl\label{lemma:convex_order_batch}\textsc{(Comparison of Busy
Periods under  Convex Order \& Poisson Arrivals).} For two
$M/GI/\infty$ queues $Q_k,Q_m$ with i.i.d Poisson arrivals with
rates $\lambda_k$, $\lambda_m$ and i.i.d  service times $T_k,T_{m}$,
we have \beq \lambda_k T_k \lcx \lambda_m T_m \Rightarrow  B_k
\leqst   B_{m}.\label{eqn:lemmabatchcx}\eeq \el

\bprf See~\ref{proof:convex_order_batch}. \eprf

Informally, the above result      states that a   more variable
service distribution (normalized by the arrival rate) results in
larger busy periods.

The results in \eqref{eqn:lemmabatchstuncond} and
\eqref{eqn:lemmabatchcx}  form an integral component of our proofs
in the comparison of tracking accuracies since, larger busy periods
leads to  lower tracking accuracies. However, we   see in the
subsequent sections that   certain additional conditions, in
addition to stochastic or convex orders of arrivals and services,
are   needed to guarantee the order of the tracking accuracies, and
hence, optimality of our heuristics for instrumentation allocation.

\section{Optimality in $GI/GI/\infty$ Queues}\label{sec:comp}

\subsection{Load-Factor Heuristic}\label{sec:load_comp}

Recall that the load-factor heuristic, described in Section
\ref{sec:loadheuristic},  predicts queues with higher load factors
to have lower timestamp-based tracking  accuracy and hence,  selects
them for introducing identifiers through instrumentation. We now
provide sufficient conditions on the arrival and service processes
under which  the load-factor heuristic is the optimal resource
allocation strategy.

A stochastic order on the   arrival and the service times is a
prerequisite condition in our approach since it leads to a
stochastic order on the busy periods from
Lemma~\ref{lemma:busyperiod}. In addition to the stochastic order on
the arrival and service processes, we need   additional  conditions
to establish the order of the tracking accuracies, depending on the
matching policy employed.  These additional conditions turn out to
be   different for the FIFO and the random matching rule. This is
because the tracking accuracies of the two rules are sensitive to
different kind of events.  For  the FIFO rule, any out-of-order
departure results in an error, which implies its sensitivity to the
\emph{spread} of the service distribution, defined precisely in
Section~\ref{sec:fifocomp}. On the other hand, random matching is
somewhat less sensitive to the service spread since it uniformly
picks a matching out of all valid matchings, and this is reflected
in our results.  We first provide sufficient conditions for
optimality of the load-factor heuristic under the FIFO rule and then
consider the random matching rule. Finally, in
Section~\ref{sec:scale}, we provide  examples where these conditions
are satisfied.

\subsubsection{Optimality Under  FIFO Matching
Rule}\label{sec:fifocomp}

We now provide conditions for the optimality of the load-factor
heuristic when FIFO is the matching policy employed at all the
queues. Since overtaking or out-of-order departures cause   errors
in FIFO matching, we relate  the tendency for overtaking to the
\emph{spread} of the service distribution, given by
\beq\label{eqn:spread_def}\spread_k
\defeq  T_k(1)-T_k(2)  ,\eeq where $T_k(1)$ and $T_k(2)$ are
independent samples of the service time   $T_k$ at queue $Q_k$. Note
that $\spread_k \equiv 0$, if the service  is deterministic. The
spread of a distribution is thus related to the variability;  a more
``spread out" service distribution has higher variability, and thus,
has higher tendency for generating out-of-order departures.

We now show the main result   that the order of the tracking
accuracies under  FIFO rule follow the reverse order of the load
factors in the presence of a stochastic order.

\begin{theorem}\label{thm:fifo}\textsc{(Optimality of Load-Factor Heuristic Under FIFO
Rule).}  At queues $Q_k, Q_m$, under a  stochastic order  on arrival
times $X_k$ and $X_m$, service times $T_k$ and $T_m$ and their
spreads $V_k$ and $V_m$,  we have
\begin{align}\nn& X_k \leqst X_m,\,
 T_k\geqst   T_{m},\,|\spread_k|\geqst   |\spread_{m}|\\ &\Rightarrow \rho_k\geq
\rho_m, \,  P^\fifo(k)\leq
 P^\fifo(m)\label{eqn:fifocond}.\end{align} Hence, if the arrival, service and service
 spread distributions  at all the queues  satisfy
the above stochastic order, then the load-factor heuristic for
allocation of instrumentation resources  is optimal, according to
optimization in \eqref{eqn:onestepopt}.  \et

\bprf  See~\ref{proof:fifo}.  \eprf

Hence,  slower arrivals, faster services (which thus imply a lower
load factor), and lower service spreads result  in more accurate
tracking  under the FIFO rule, when the comparison is formalized by
the notion of stochastic order.

The combined conditions of service speed and spread in
\eqref{eqn:fifocond} places constraints on the service distributions
under comparison. Informally, we need one service to be
simultaneously slower and more spread out than the other, \ie one
service distribution has   more  probability mass concentrated
closer  to zero   than the other. For example, the  Weibull
distribution with different shape   parameters but same  scale
parameter satisfies this condition, as shown in
Fig.\ref{fig:weibull}.


\begin{figure}[t]\bc
\bp\psfrag{shape=2}[l]{\scriptsize $w=2$}
\psfrag{shape=4}[l]{\scriptsize $w=8$}
\includegraphics[width=2.2in]{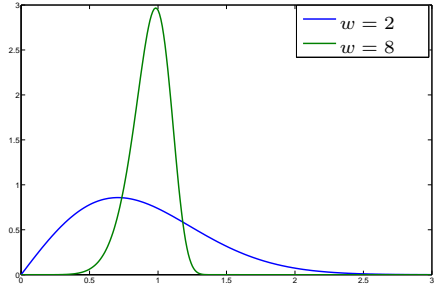}\ep
\ec\caption{Comparison of two Weibull distributions with  pdf
$f(x)=(\frac{w}{c}) (\frac{x}{c})^{w-1} \exp(-(\frac{x}{c})^w)$ for
$x>0$. The distribution with lower shape parameter $w$ has higher
FIFO tracking accuracy. See Theorem
\ref{thm:fifo}.}\label{fig:weibull}
\end{figure}

\subsubsection{Optimality Under Random Matching}\label{sec:randcomp}

We now provide sufficient conditions for optimality of the
load-factor heuristic when the random matching rule is employed for
matching arrival and departure timestamps at all the queues. Recall
that   random matching rule  uniformly chooses a  matching among all
valid matchings in the busy period.

We now show the main result of this section that the order of the
tracking accuracies under  the random matching rule follow the
reverse order of the load factors in the presence of a stochastic
order.

\bt\label{thm:rand}\textsc{(Optimality of Load-Factor Heuristic
Under Random Matching Rule).}  At queues $Q_k, Q_m$, under  random
matching rule with  arrival times $X_k$ and $X_m$, service times
$T_k$ and $T_{m}$ with supports, $\supp(f_{T_k})=
[\alpha_k,\beta_k]$ and $\supp(f_{T_m})=[\alpha_m,\beta_m]$, we have
\begin{align}\nn &X_k \leqst X_m, \, T_k \geqst   T_{m},\, \alpha_k\leq
\alpha_m \\ &\Rightarrow \rho_k\geq \rho_m, \,   P^\rand(k)\leq
 P^\rand(m).\label{eqn:cond_rand} \end{align} Hence, if the arrival, service and service
 support  at all the queues  satisfy
the above stochastic order, then the load-factor heuristic for
allocation of instrumentation resources  is optimal, according to
optimization in \eqref{eqn:onestepopt}.  \et

\bprf  See~\ref{proof:rand}. \eprf

Hence, slower arrivals and faster services along with a mild
condition on the support lower bounds of the service distribution
imply lower tracking accuracy under the random matching rule, when
the comparison is formalized by a stochastic order.

The condition in \eqref{eqn:cond_rand} on the support of the service
distributions   is  mild and is usually satisfied since one mostly
encounters service distributions with  a lower  bound   of support
equal to zero. However, it   cannot be dropped as seen in this
example when $T_k \equiv \mu_k$ and       $T_{m}=\unif(0,2\mu_m)$,
the uniform distribution, with $\mu_k> \mu_m$. Since $\alpha_k =
\mu_k > \alpha_m=0$, \eqref{eqn:cond_rand} does not hold, which is
indeed true since in fact, $P^\rand(k)=1>
 P^\rand(m)$  in this example.

\subsubsection{Special Case: Same Distribution
Family}\label{sec:scale}

We have so far established sufficient conditions for optimality of
the load-factor heuristic when all the queues employ either the FIFO
or the random matching rules.   We now consider a special case of
arrival and service distributions belonging to the same distribution
family  where optimality of the load-factor heuristic is guaranteed
under both FIFO or random matching rules, without the need for
additional conditions.

\begin{corollary}\label{cor:rescale}
\textsc{(Optimality of Load-factor Heuristic Under Same Distribution
Family).} When   the  service distributions at different queues are
linearly scaled versions of the same distribution, and the same
holds for all the arrival distributions as well, then the  tracking
accuracies at the queues are in the reverse order of their load
factors under both FIFO and random matching rules. Hence, here, the
load factor heuristic is optimal for resource allocation, according
to optimization in \eqref{eqn:onestepopt}.
\end{corollary}

\bprf We show that the conditions for FIFO rule   in  Theorem
\ref{thm:fifo} are satisfied. For random matching rule, the
condition on lower bound  of support in Theorem \ref{thm:rand} is,
however, violated. Hence, we need to prove the above statement from
scratch. See~\ref{proof:rescale}. \eprf

The above result holds if all the service distributions are say
exponential, uniform  and so on.  In practice, the service
distributions of different subsystems may be similar and hence, this
result may be relevant. The constraint on the arrival processes is
however more restrictive in case of an inter-connected network of
queues,  since it limits to  Poisson arrivals to the system.

\subsection{Unit-Batch Approximation}\label{sec:load_lr}

We have so far demonstrated the effectiveness of the load-factor
heuristic  when the arrival and service distributions are similar or
more generally, constrained to satisfy a stochastic order.  Next, we
provide sufficient conditions to establish the optimality of  the
alternative heuristic for instrumentation allocation based on
unit-batch approximations, described in Section~\ref{sec:eval}.
Recall that the unit-batch approximation selects queues for
instrumentation  in the increasing  order of their probability of
having a unit-sized busy period.

\subsubsection{Optimality Under Stochastic Order}

We now show that the conditions given in Theorems~\ref{thm:fifo} and
\ref{thm:rand}, which guarantee optimality of the load-factor
heuristic, also guarantee the optimality of the unit-batch
approximation.

\bt\textsc{(Optimality of Unit-Batch Approximation Under Stochastic
Order).} We have for two queues $Q_k$ and $Q_m$, \begin{align}\nn&
X_k \leqst X_m,\,
 T_k\geqst   T_{m},\,|\spread_k|\geqst   |\spread_{m}|\\ &\Rightarrow \Pbb[B_k=1]\leq \Pbb[B_m=1], \,  P^\fifo(k)\leq
 P^\fifo(m)\label{eqn:fifocondapprox}.\end{align}
 \begin{align}\nn &X_k \leqst X_m, \, T_k \geqst   T_{m},\, \alpha_k\leq
\alpha_m \\ &\Rightarrow \Pbb[B_k=1]\leq \Pbb[B_m=1], \,
P^\rand(k)\leq
 P^\rand(m).\label{eqn:cond_randapprox}\end{align} Hence, the above conditions
guarantee that the heuristic based on unit-batch approximation
coincides with the load-factor heuristic and hence, also achieves
optimality in \eqref{eqn:onestepopt}.
 \et

\bprf It is easy to see that $\Pbb[B_k=1]= \Pbb[X_k> T_k]\leq
\Pbb[B_m=1]$ since $X_k - T_k \leqst X_m - T_m$. \eprf

Hence, the unit-batch approximation achieves optimality in the above
scenario where the load-factor heuristic is also optimal. We now
demonstrate the superiority of the unit-batch approximation over the
load-factor heuristic by considering a different scenario.

\subsubsection{Optimality Under Convex Order}

We now consider a special scenario where all the queues have the
same load factor but with different service variabilities. In this
case, the load-factor heuristic fails to distinguish the tracking
accuracies of different queues and its performance is equivalent to
a random selection of queues for instrumentation. On the other hand,
we  show below that the unit batch approximation  achieves optimality
when the queueing services satisfy a convex order.

\bt\label{thm:convex_order_opt}\textsc{(Optimality of Unit-Batch
Approx. Under Convex Order and FIFO Rule).}   For two $M/GI/\infty$
queues $Q_k,Q_m$ with i.i.d Poisson arrivals with rates $\lambda_k$,
$\lambda_m$ and i.i.d  service times $T_k,T_{m}$,  we
have\begin{align}\nn &\lambda_k T_k \lcx \lambda_m T_{m}\\ &
\Rightarrow \Pbb[B_k=1] \geq \Pbb[B_{m}=1],\,P^\fifo(k) \geq
 P^\fifo(m).\label{eqn:convex_order_opt}\end{align}  Hence, under Poisson arrivals,
 convex order of normalized services and FIFO matching rule,  the unit-batch approximation
is the optimal
 strategy for
allocation of instrumentation resources, according to optimization
in \eqref{eqn:onestepopt}. \et

\bprf $\Pbb[B_k=1] = \Ebb[e^{-\lambda_k T_k}] $ is a concave
function in $\lambda_k T_k$ and hence, $\Pbb[B_k=1]\geq
\Pbb[B_m=1]$. For the order of $P^\fifo(k)$ and $P^\fifo(m)$, see~\ref{proof:convex_order_opt}.\eprf


Hence, the unit-batch approximation achieves optimality over a wider
range of distributions than the load factor heuristic. The relative
performance of the load-factor heuristic and unit-batch
approximation for instrumentation allocation depends on the queues
under consideration. For queues with similar service distributions
but significantly different load factors, the load-factor heuristic
suffices to achieve  efficient   allocation.  On the other hand, if
all the load factors are close to one another, the effect of service
variability and   higher-order moments become significant and are
not captured by the load-factor heuristic. In such scenarios, there
is significant advantage in employing the unit-batch approximation.

\section{Product-Form Networks}\label{sec:bcmp}

We have so far considered comparison of monitoring performance for
different service distributions when all the queues are
infinite-server queues. In this section, we extend some of our
results to the more general queueing networks consisting of
egalitarian processor sharing (PS) queues (with load factors less
than one to ensure stability) and the infinite-server queues.  These
are part of the well-known product-form queues\footnote{\scriptsize
The tracking accuracy of $GI/M/1$ with first-come first-serve (FCFS)
or last-come first-serve with preemption (LCFS-PR), which are part
of a product-form network,  is unity. This is because  there is a
fixed order of departures. Hence, they are ignored for
instrumentation   allocation.} \cite{Baskett&etal:75JACM}.

\subsection{Processor-Sharing   Network}

We first consider  all the queues  to be processor-sharing queues
which makes comparison between them tractable. In the (egalitarian)
processor sharing, each waiting transaction gets an equal share of
service capacity.  Since there is simultaneous processing of
transactions, out-of-order departures are possible and  there is
uncertainty in   matching arrival and departure timestamps.

In a nutshell, we now show that the comparison results for
infinite-server queues under random matching in
Theorem~\ref{thm:rand} holds for processor-sharing queues as well.
However, the proof is more involved since the sojourn time
distributions of different transactions are correlated under
processor-sharing discipline.

We use the term {\em job-length}  to refer to the amount of service
required, and we use the term {\em sojourn time} to denote the
amount of time spent in the system. We  denote the   job-lengths by
$\bfJ=[J(1),J(2),\ldots]$ and assume that $J(i)
\overset{i.i.d.}{\sim} f_J$.

\bt\label{thm:ps} \textsc{(Optimality of  Heuristics   in
Processor-Sharing Queues Under Random Matching).} Given two
processor-sharing queues with  job lengths $J_k$ and $J_{m}$  and
supports $[\alpha_k,\beta_k]$ and $[\alpha_m,\beta_m]$, we have
\begin{align}
\label{eqn:ps_rand} & J_{k} \geqst J_{m}, \alpha_k\leq \alpha_m, \\
\nn &\Rightarrow  \rho_m\geq \rho_k, \, \Pbb[B_m=1]\leq
\Pbb[B_k=1],\,  P^\rand(k)\leq
 P^\rand(m).\end{align} \et

\bprf See~\ref{proof:ps}.  \eprf

The above results on comparison of two processor-sharing queues
under  random matching are identical to those comparing two
infinite-server queues in Theorem~\ref{thm:rand}. Hence, our
heuristics are optimal for instrumentation under the above
stochastic-order conditions when all the queues are either
infinite-server or processor-sharing queues. However, when we have
both infinite-server and processor-sharing queues, the above results
are no longer valid and we consider this scenario in the next
section.

\subsection{Product-Form Network}

We now compare monitoring performance of a processor-sharing queue
with an infinite-server queue. This analysis is   more complicated
since the sojourn times of the two queues have   different
dependency structures. We  limit to the scenario when the job
lengths in the processor-sharing queue  stochastically dominate the
service times  of the infinite-server queue.

\bt\label{thm:inf_ps}\textsc{(Optimality in Product-Form Networks
Under Random Matching).}Given a processor-sharing queue with
job-lengths $  J_\psq$  with    support $[\alpha_\psq,\beta_\psq]$
and infinite-server queue with service $T_\inft$, and arrivals
$X_\psq$ and $X_\inft$,  \begin{align}\label{eqn:inf_ps}&
X_\psq\leqst X_\inft,\,  J_\psq \geqst T_\inft, \,\alpha_\psq\leq
\alpha_\inft\\ & \Rightarrow \rho_\psq\geq \rho_\inft,
\,\Pbb[B_\psq=1]\leq \Pbb[B_\inft=1],\,P^\rand_\psq \leq
 P^\rand_\inft.\nn\end{align}  \et

\bprf See~\ref{proof:inf_ps}\eprf

Hence, in a product-form network, under the above stochastic order,
our two heuristics coincide with the optimal instrumentation
strategy.

\section{Numerical Analysis}\label{sec:sim}

We have so far provided a precise set of theoretical conditions when
the two proposed heuristics coincide with the optimal
instrumentation allocation strategy. In this section, we compare the
performance of various instrumentation strategies through
simulations. There are mainly two questions we seek to answer: How
do our heuristics compare with the optimal solution when the
theoretical conditions in Sections~\ref{sec:comp} and \ref{sec:bcmp}
for optimality are not met?  What is the relative performance of the
two heuristics in different load regimes?

We   consider infinite-server queues with service distributions
belonging to the  Weibull family.  The Weibull distribution is a
rich family   allowing us to tune the rate and the randomness of the
service time separately by varying the scale and the shape
parameters, and also includes the exponential distribution (shape
parameter $w=1$). Note that for the same scale parameter $c$, the
variance decreases with the shape parameter $w$. Hence,
distributions with $w<1$ have higher variance than the exponential
distribution, and vice versa.

\begin{figure*}[t]
\centerline{\small Arrival rate $1$ of Poisson process, $1000$
transactions, $10$ Monte Carlo runs.} \subfloat[a][Shape Parameter
$w=1$]{ \label{fig:singlelink_Exp}
 \begin{minipage}{2.1in}
 \begin{center}
\begin{psfrags}\psfrag{ML}[l]{\scriptsize ML}\psfrag{FIFO}[l]{\scriptsize FIFO}
\psfrag{rand}[l]{\scriptsize Rand}\psfrag{low-rate}[l]{\scriptsize
Unit batch} \psfrag{service rate}[c]{\small Service Rate}
\psfrag{batch matching probability}[l]{\scriptsize }
\includegraphics[width
=2.1in]{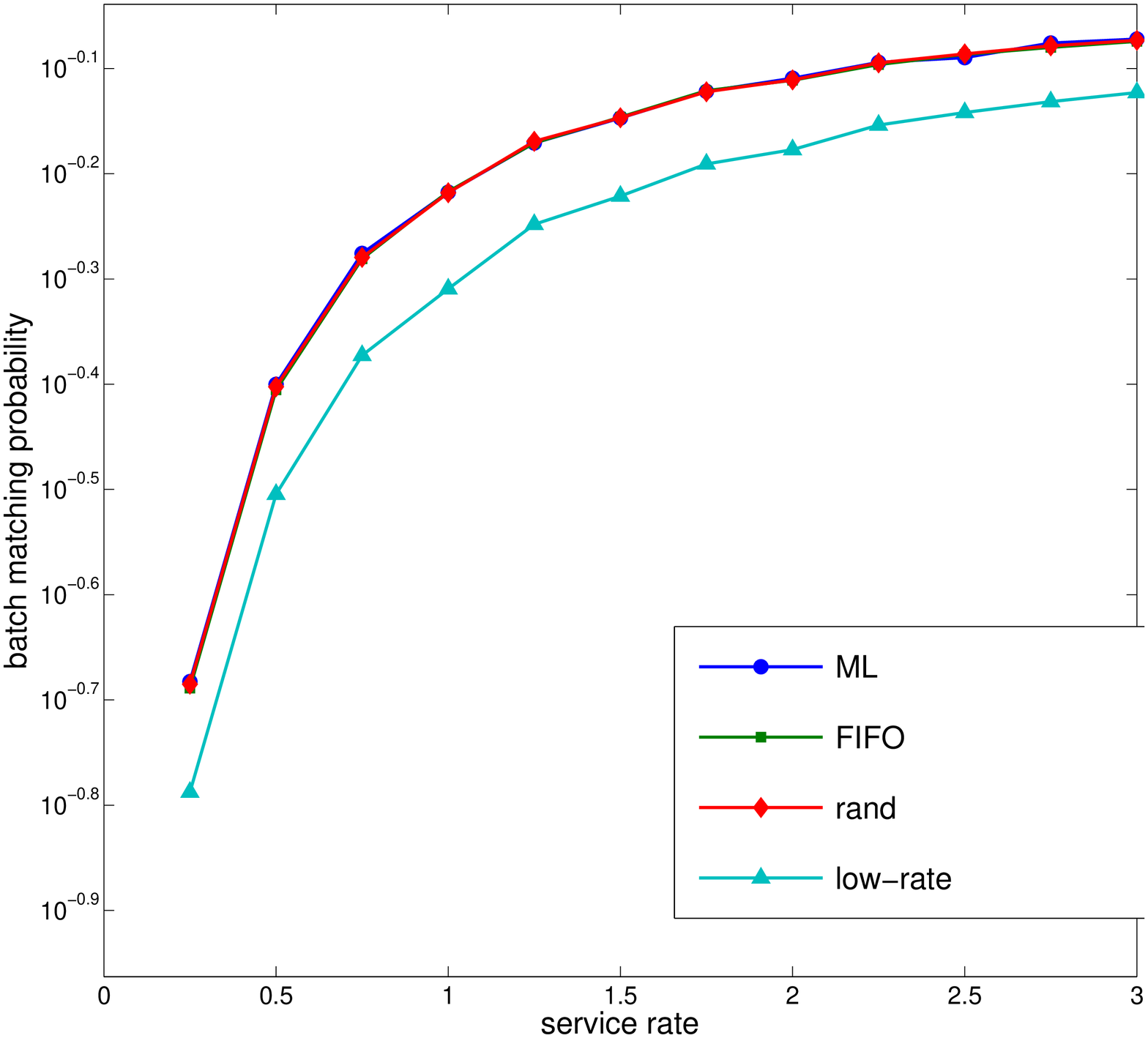}
\end{psfrags}
\end{center}
 \end{minipage}}\hfil\subfloat[b][Shape Parameter $w=1.5$]{ \label{fig:singlelink_heavyWeibull}
 \begin{minipage}{2.1in}
 \begin{center}
\begin{psfrags}\psfrag{ML}[l]{\scriptsize ML}\psfrag{FIFO}[l]{\scriptsize FIFO}
\psfrag{rand}[l]{\scriptsize Rand}\psfrag{low-rate}[l]{\scriptsize
Unit batch}\psfrag{service rate}[c]{\small Service Rate}
\psfrag{batch matching probability}[l]{\scriptsize }
\includegraphics[width
=2.1in]{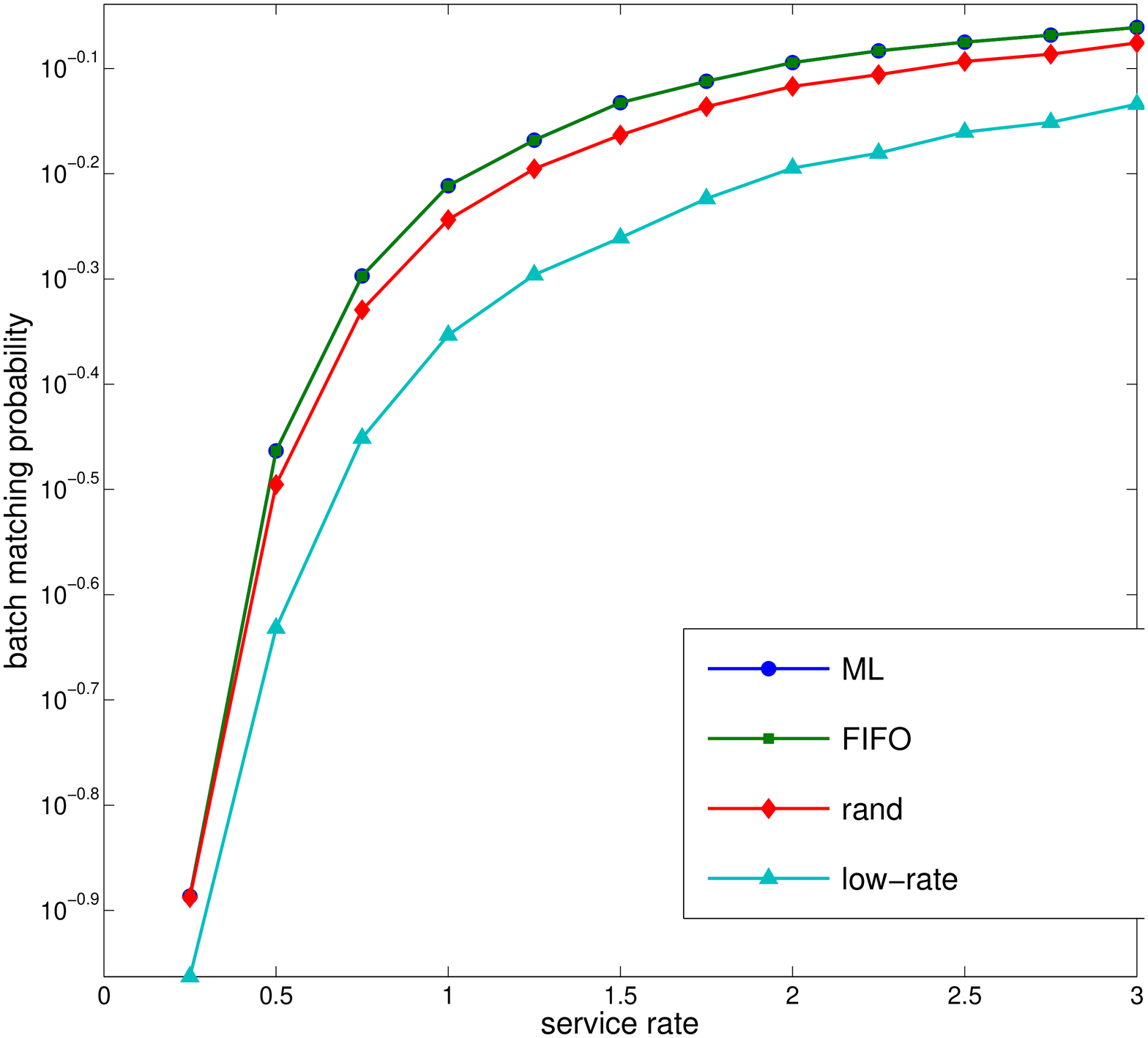}
\end{psfrags}
\end{center}
 \end{minipage}}\hfil\subfloat[c][Shape Parameter $w=0.5$]{ \label{fig:singlelink_lightWeibull}
 \begin{minipage}{2.1in}
 \begin{center}
\begin{psfrags}
\psfrag{ML}[l]{\scriptsize ML}\psfrag{FIFO}[l]{\scriptsize FIFO}
\psfrag{rand}[l]{\scriptsize Rand}\psfrag{low-rate}[l]{\scriptsize
Unit batch} \psfrag{service rate}[c]{\small Service Rate}
\psfrag{batch matching probability}[l]{\scriptsize }
\includegraphics[width
=2.1in]{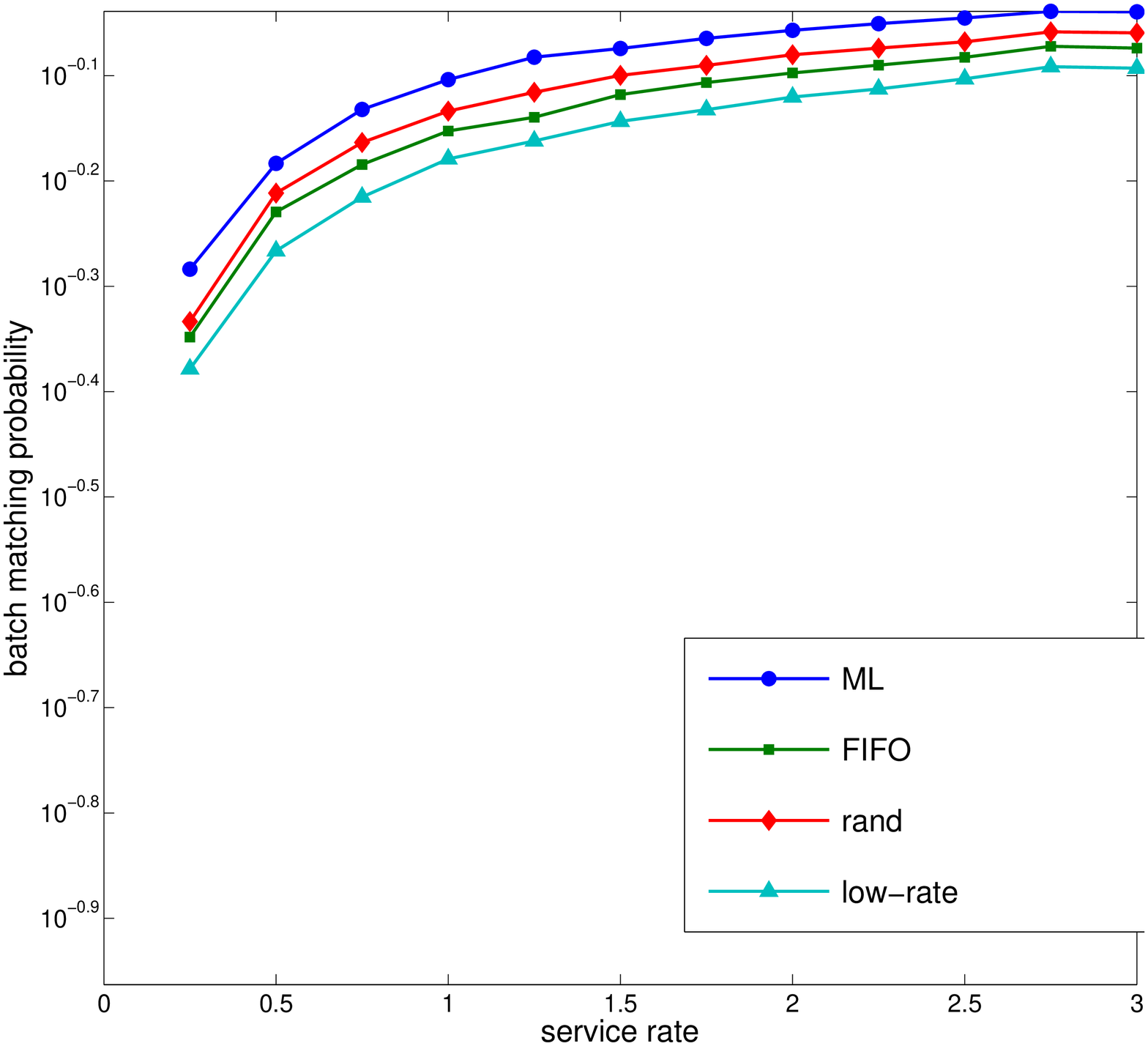}
\end{psfrags}
\end{center}
 \end{minipage}}
 \caption{Matching Accuracies of ML, FIFO, and random matching
together with the low-arrival rate approximation. See Sections
\ref{sec:match} and
\ref{sec:eval}.}\label{fig:singlelink}\end{figure*}

\begin{figure*}[t]{\small Instrument
$E=2$ out of $|\Lc|=10$ states, unit arrival rate $(\lambda=1)$ of
Poisson process, service rates $\mu_k\overset{i.i.d.}{\sim}
\unif[0.5, T_{\max}]$, Weibull shape parameter
$w_k\overset{i.i.d.}{\sim}\unif[0.1,2]$, $1000$  configurations.}

\subfloat[a][
Obj. Value]{\label{fig:multilink_value_randk}
\begin{minipage}{2.1in}
\centerline{
\begin{psfrags}
\psfrag{optimal}[l]{\tiny Optimal}\psfrag{load}[l]{\tiny Load
Factor}\psfrag{low-rate}[l]{\tiny Unit Batch}\psfrag{rand}[l]{\tiny
Rand. Select} \psfrag{mu-max}[c]{\scriptsize  Max. Service Rate
$T_{\max}$} \psfrag{value of objective}[c]{\scriptsize }
\includegraphics[width=2.1in]{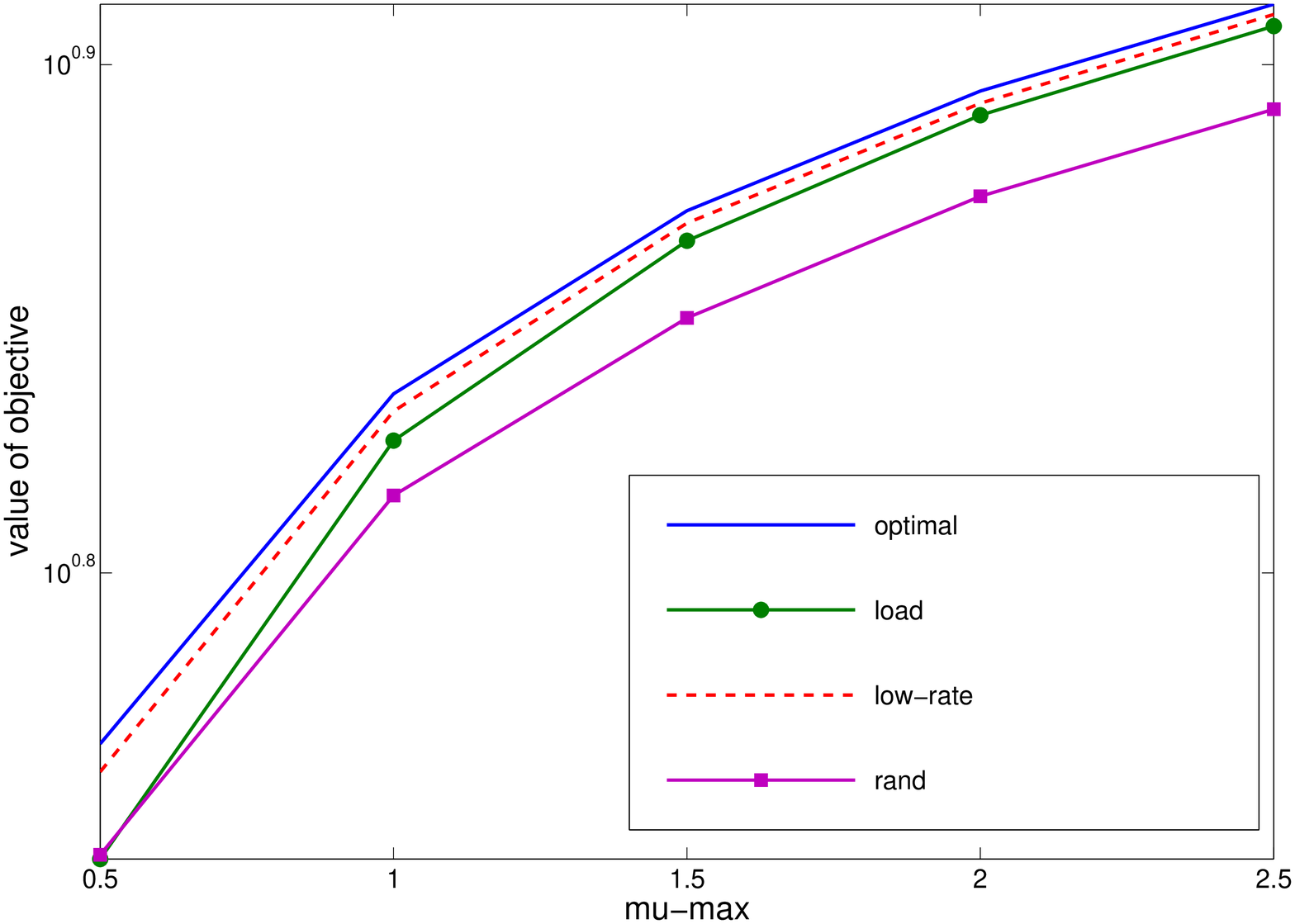}
\end{psfrags}
}\end{minipage} }\hfil\subfloat[b][Ratio = $\dfrac{\mbox{Obj. under
heuristic}}{\mbox{Optimal Obj.}}$]
{\label{fig:multilink_ratio_randk}
\begin{minipage}{2.1in}
\centerline{
\begin{psfrags}
\psfrag{optimal}[l]{\tiny Optimal}\psfrag{load}[l]{\tiny Load
Factor}\psfrag{low-rate}[l]{\tiny Unit Batch}\psfrag{rand}[l]{\tiny
Rand. Select} \psfrag{mu-max}[c]{\scriptsize  Max. Service Rate
$T_{\max}$} \psfrag{value of objective}[c]{\scriptsize }
\includegraphics[width=2.1in]{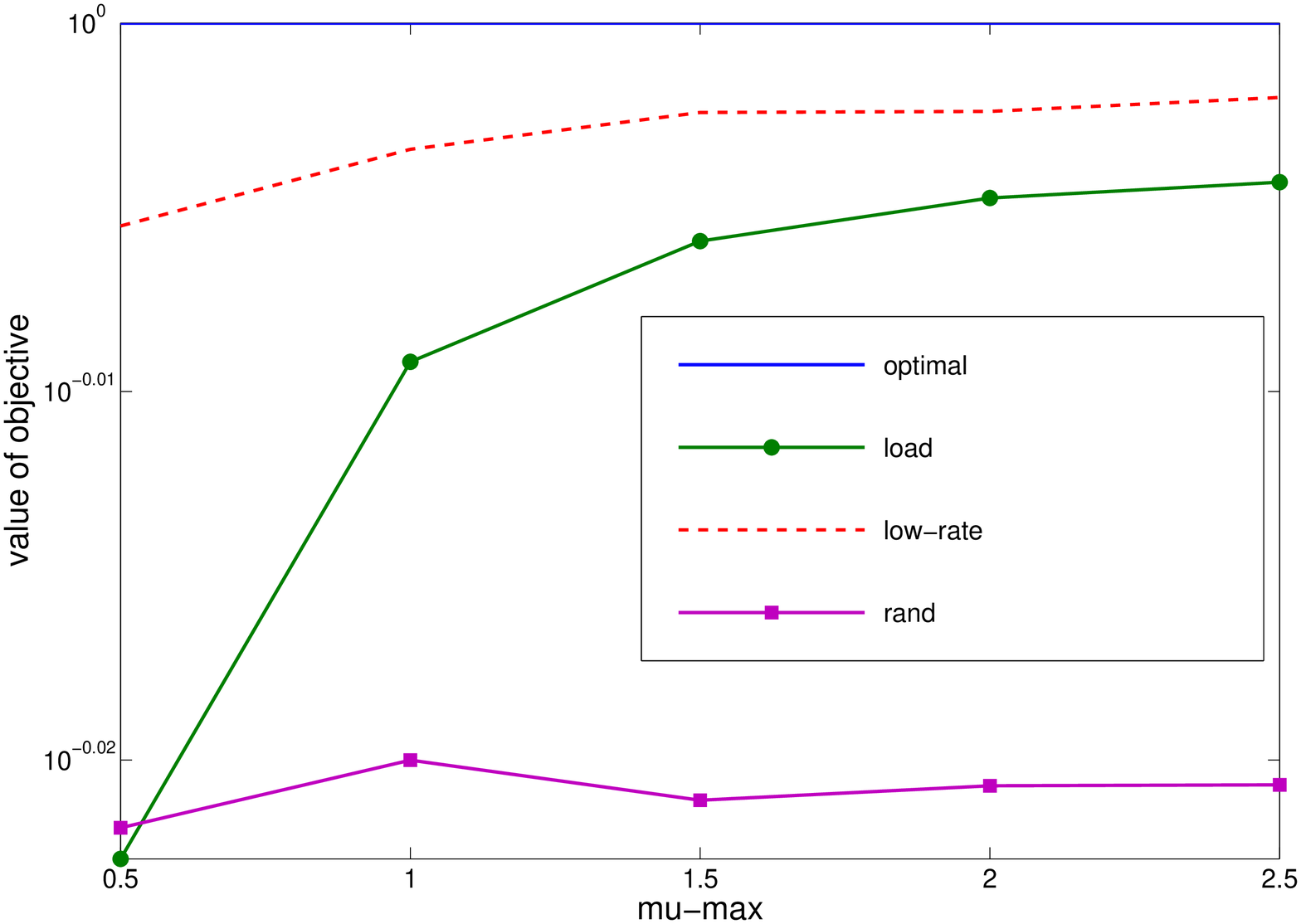}
\end{psfrags}
}\end{minipage} }\hfil\subfloat[c][Fraction of Overlap with
Opt.]{\label{fig:overlap}
\begin{minipage}{2.1in} \centerline{
\begin{psfrags}\psfrag{optimal}[l]{\tiny
Optimal}\psfrag{load}[l]{\tiny Load
Factor}\psfrag{low-rate}[l]{\tiny Unit Batch}\psfrag{rand}[l]{\tiny
Rand. Select} \psfrag{mu-max}[c]{\scriptsize  Max. Service Rate
$T_{\max}$} \psfrag{fraction of overlap}[c]{\scriptsize }
\includegraphics[width=2.1in]{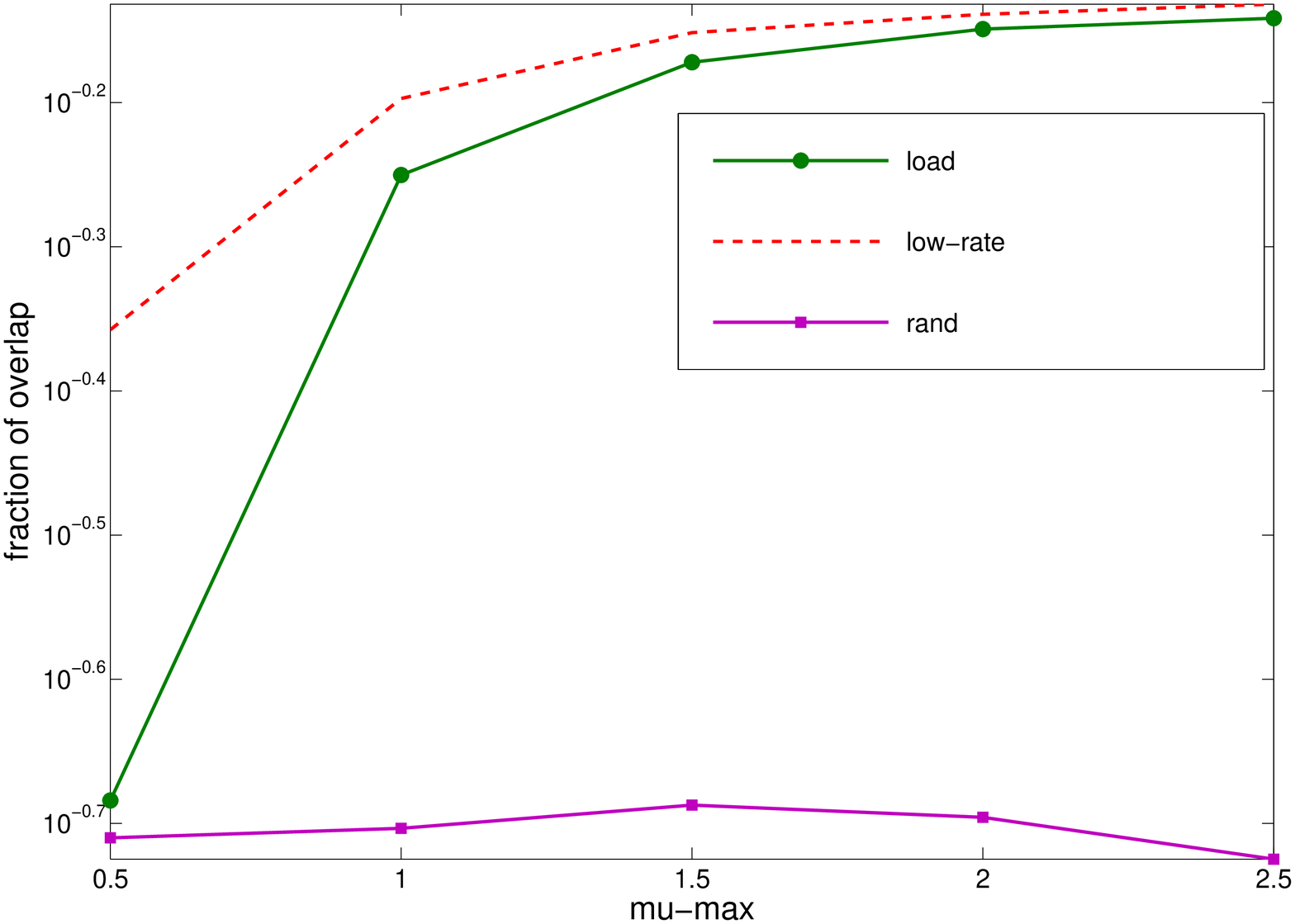}
\end{psfrags}
}\end{minipage}}\hfil \caption{Comparison of instrumentation
strategies. Obj.  $=E+\sum\limits_{Q_k \in \Lc}(1-z_k) P^{\fifo}(k)$
, see \eqref{eqn:onestepopt}.}\label{fig:multilink}
\end{figure*}

\subsection{Effect of Matching Policies}\label{subsec:single link}

In Fig.\ref{fig:singlelink}, we compare the tracking accuracies  $
P^\policy $   of  policies $\gamma$ given by the FIFO, random
matching and  the optimal maximum-likelihood (ML)   policies. We
also compare the unit-batch approximation with the exact tracking
accuracy. In Fig.\ref{fig:singlelink_Exp}, for the shape parameter
$w=1$, we have the exponential distribution, and all the matching
policies, \viz ML, random, and FIFO matchings have equal
performance,  consistent with the analytical results in
\cite{Anandkumar&Bisdikian&Agrawal:08Sigmetrics}. In
Fig.\ref{fig:singlelink_heavyWeibull}, for  the shape parameter
$w>1$, FIFO has the same performance as ML, and is better than
random matching, again consistent with theory in
\cite{Anandkumar&Bisdikian&Agrawal:08Sigmetrics}\footnote{
It is shown in \cite{Anandkumar&Bisdikian&Agrawal:08Sigmetrics} that FIFO matching coincides with the optimal ML tracking when the shape parameter $w>1$, i.e, there is less variation in service times.}. In
Fig.\ref{fig:singlelink_lightWeibull}, for the shape parameter
$w<1$, we have heavy-tailed services, and here, random matching has
better accuracy than   FIFO  rule. This is intuitive since out-of-order departures are more likely under heavy-tailed services. Moreover, the
tracking accuracy in all these cases increases with the service rate
as predicted.

In all the cases, there is a non-trivial gap between the actual
tracking accuracies and the unit-batch  approximation (up to about
$10\%$); however, the approximation correctly follows the trend of
the true values. Hence, we can expect  solutions based on the exact
and approximate evaluation  to pick a similar set of queues for
instrumentation, thereby leading to efficient allocation of
monitoring resources, as discussed below.

\subsection{Comparison of Instrumentation Strategies}\label{subsec:multilink}

In Fig.\ref{fig:multilink}, we   compare   our instrumentation
strategies based on the load factor   and the unit-batch
approximation    with   the optimal strategy under the optimization
rule in (\ref{eqn:onestepopt}). As a benchmark, we also  compare the
proposed strategies with random instrumentation, \ie uniformly
selecting a subset of queues for instrumentation.

We consider   Weibull service times and FIFO matching (similar
results are observed under random matching). We run simulations
under randomly chosen parameters  for each queue and then average
the results of different configurations. Specifically, the service
rates are drawn i.i.d. uniformly between a minimum and a maximum
service rate, and so are the  shape parameters. We vary the maximum
service rate to obtain more diverse set of service distributions for
the queues under consideration for instrumentation allocation. Since
the parameters are randomly chosen, the sufficient conditions for
optimality of our heuristics proven in Section~\ref{sec:comp} are
not met, and we do not expect our heuristics to exactly coincide
with the optimal allocation strategy.


In Fig.\ref{fig:multilink}, we see that the performance of the two
heuristics gets closer to that of the optimal strategy as the
maximum service rate increases leading to a more diverse set of
queues. For the load-factor heuristic, this is because the load
factors of different queues are well separated as the queues become
more diverse. For the unit-batch heuristic, this is because, in
addition, the service rates are increasing on average, leading to
tighter approximation of the tracking accuracy. On the other hand,
the gap between  optimal allocation and random allocation increases
with the maximum service rates since random allocation performs
poorly when the queues are diverse.  We also note that the
performance of the unit-batch approximation is superior over the
load-factor heuristic but they become close when the queues have
well-separated load factors, as predicted in
Section~\ref{sec:2approaches}.

\section{Conclusion}\label{sec:conc}

In this paper, we considered the problem of optimal instrumentation
allocation for tracking transactions in a queueing network. Two
types of monitoring resources are considered in the form of
identifiers and timestamps. Identifiers provide precise tracking but
are limited while timestamps are imprecise but available everywhere.
The optimal allocation strategy selects queues with least
timestamp-based tracking accuracies for introducing identifiers. We
proposed two simple heuristics for allocation which coincides with
the optimal strategy under certain conditions on  arrival and
service processes. Simulations show that our solutions are effective
even when there is a deviation from the optimality conditions.

While providing a strong theoretical foundation and effective
solutions for instrumentation allocation, we acknowledge that the
overall problem has a broader  range of challenges. For instance, in
practice, the model for arrivals and services may not be known and
needs to be estimated from data as well. There may be systems where
complete timestamp information may not be available. We have assumed
equal costs for instrumenting different components, while  with
unequal costs, we need to investigate new optimality conditions for
our heuristics.  We have assumed an infinite-server queueing system, while in reality there are a finite number of servers. The optimality results can in principle be extended to this scenario.  However, direct analysis of such a system is much more involved since the service times of different packets are not independent. Moreover,  the infinite-server system is the worst-case scenario for timestamp-based   tracking since a finite-server system is less likely to produce out-of-order transactions. In this sense, the recommended instrumentation solution can be viewed as maximizing a lower bound on the tracking accuracy under finite-server queueing.
Other challenges involve  analyzing the effect of
admission control and allowing for dynamic switching of data
collection between different systems.

\subsection*{Acknowledgements}
The authors thank R. Nunez Queija for discussions on the
processor-sharing queue and Varun Gupta for  discussions on the
notion of convex order at the MAMA 2009 workshop.


\begin{appendix}

\section{Accuracy Under FIFO}\label{proof:fifoexpect}

\bl\label{lemma:fifoexpect}  The tracking
accuracy in  \eqref{eqn:acc} simplifies under   FIFO rule as\[
P^\fifo =  \sum_{b=1}^\infty\Pbb[\pibf^t =  \bfI, B=b],\]  where
each term in the series  $\Pbb[\pibf^t =  \bfI,B=b]$ is given by  \[
=\Pbb(\bigcap_{i =1}^{b-1}\{ T(i) \in [X(i), X(i) + T(i+1)]\} \cap
\{T(b) < X(b)\}),\] where $X(i)$ and $T(i)$ are the inter-arrival
and service times.  \el

\bprf Given the busy-period size $B=b$, the event  that   FIFO rule
is correct is \beq\nn\label{eqn:fifoevent} \Ac_b^\fifo  =
\bigcap_{i=1}^{b-1}\{ T(i)< X(i) + T(i+1)\} ,\eeq since $i^{\tha}$
transaction needs to depart sooner than the $(i+1)^{\tha}$
transaction. The  event that the  busy-period size is $B=b$ is given
by \beq\nn \label{eqn:busyperiodevent} \{B =b\}  =
 \bigcap_{i =1}^{b-1}\{
 T(i)\in [X(i),\sum_{j =i}^b X(j)]\}\cap\{  T(b)<X(b) \} . \eeq
$P^\fifo=\sum_{b=1}^\infty \Pbb[\Ac_b^\fifo \cap \{B=b\}]  $ and
result follows. \eprf

\section{Accuracy Under Random Rule}\label{app:random} I
n order to compute the tracking accuracy $P^\rand$ under random
matching rule, we  need to find the number of valid matchings. The
number of     such valid matchings is given by the number of perfect
matchings in  the $0$-$1$ biadjacency matrix  $\bfA_k$ defined as
follows: for a bipartite graph with  arrivals $\bfY_k$ in one
bipartition and  departures $\bfD_k$ in the other, the presence of
edge $(i,j)$ in $\bfA_k$ indicates   positive likelihood of
$i^{\tha}$ arrival corresponding to the $j^{\tha}$ departure \beq
\label{eqn:amatrix_support} A_k(i,j) =1 \iff f_{T_k}[D_k(j) -
Y_k(i)]>0, \quad \forall 1\leq i,j\leq b.\eeq Any valid matching
between the arrivals and the departures is a perfect matching  on
the biadjacency matrix $\bfA_k$, where    a perfect matching    is
defined as a  set of pairwise non-adjacent edges where all vertices
are matched. The number of perfect matchings for the biadjacency
matrix $\bfA$ is given by its permanent  \beq \perm(\bfA)
\defeq\sum_{\pibf} \prod_{i=1}^b
A(i,\pi(i)),\label{eqn:perm}\eeq where the sum is over all the
permutation vectors $\pibf$ over  $\{1$ $,\ldots,$ $ b\}$
conditioned on busy period size $ B=b$. Denote the perfect matching
chosen by   random matching  as $\pibf^\rand$.    Since each perfect
matching is chosen with uniform probability and there are
$\perm(\bfA)$ number of them, the probability of choosing one of
them is $\perm(\bfA)^{-1}$.   Using this fact, it is easy to now
derive the expression for tracking accuracy under random matching
\begin{align}\nn P^\rand&=\sum_{b=1}^\infty\Pbb[\pibf^\rand=
\pibf^t, B=b],\\  \label{eqn:randexpectseries}&= \sum_{b=1}^\infty
\sum_{\bfa} \frac{\Pbb[\bfA= \bfa, B=b]}{\perm(\bfa)}. \end{align}

\section{Introduction to Stochastic Order}\label{app:stocorder}

\subsubsection{Stochastic  Order}

The stochastic order (also known as the usual stochastic order) is
defined as follows
\cite{Shaked&Shanthikumar:book,Muller&Stoyan:book}.

\bd[Stochastic Order] A variable $Z_1$ is said to be stochastically
dominant with respect to a variable  $Z_2$, denoted by $Z_1\geqst
Z_2$, if\beq \label{eqn:stoc_comp_inc} Z_1\geqst  Z_2 \iff
\Ebb[\phi(Z_1)]\geq \Ebb[\phi(Z_2)],\eeq for all increasing
functions $\phi$ for which   expectations exist. \ed

Naturally, the above definition   implies
\beq\label{eqn:stoc_imp_means} Z_1\geqst   Z_2 \Rightarrow
\Ebb[Z_1]\geq \Ebb[Z_2].\eeq 

We intend to compare tracking accuracies at queues when their
arrival processes satisfy a certain stochastic order and their
service processes satisfy the reverse stochastic order. We  leverage
on the stochastic orders to guarantee an order on the tracking
accuracies at different queues and hence, optimality of our
heuristics.

\subsubsection{Convex Order}

We define another notion of comparison of random variables known as
the {\em convex order}  \cite[Ch. 3]{Shaked&Shanthikumar:book}.

\bd[Convex Order]A variable $Z_1$  is said to be smaller than $Z_2$,
denoted by $Z_1\lcx Z_2$, if for all convex functions $\phi :\Re
\mapsto \Re$, $\Ebb[\phi(Z_1)]\leq \Ebb[\phi(Z_2)]$.\ed

%


The convex order compares the variability of  random variables and
requires equal mean values,
\[Z_1\lcx Z_2 \Rightarrow \Ebb[Z_1]=\Ebb[Z_2], \Var[Z_1] \leq \Var[Z_2].\]
In our context, we intend to compare queues under the same load
factor but with different variability in services. Intuitively, a
service distribution with higher variability results in more
uncertainty in the order of departures   implying lower tracking
accuracy, and we use the notion of convex order to capture this
effect.


The stochastic and convex orders thus deal with different aspects of
comparison of  random variables: the former deals with the
magnitudes while the latter deals with variability, and one does not
imply the other. There are many sufficient conditions which can be
easily checked  for the stochastic or convex order to hold
\cite{Shaked&Shanthikumar:book}. For a set of queues, we can use
these conditions to check if the stochastic or the convex orders
hold, in which case, we can draw conclusions about the optimality of
our heuristics for instrumentation allocation.
\section{Proof of Lemma
\ref{lemma:busyperiod}}\label{proof:busyperiod}

We have for $b\geq 1$, \[ \Pbb[B_k=b]  =  \Pbb[\!\!\!\!\bigcap_{i
=1, \ldots, b-1}\!\!\!\!X_k(i) \leq T_k(i)\leq \sum_{j =i}^b X_k(j),
X_k(b)> T_k(b)].
\] We have $\Pbb[B_k>1] = \bar{F}_{T_k}[X_k]$ and hence,
$\Pbb[B_k>1]\geq \Pbb[B_{m}>1].$ Now consider, \begin{align} \nn
p_k(x)&\defeq\Pbb[B_k> b+1|B_k>b, X_k(b+1)=x]  \\ &=
\Pbb[T_k(b+1)\bigcup_{i=1}^b \{T_k(i) - \sum_{j =i}^b X_k(b)\}> x],
\nn
\\&= \Pbb[\max\{T_k(b+1), T_k(b)-X_k(b),\ldots,\} >x].
 \label{eqn:max} \end{align} We now claim that for $b\geq1$,
 \beq   X_k\leqst X_m, T_k\geqst
 T_{m} \Rightarrow p_k(x)\geq p_m(x) .\label{eqn:condbatch}\eeq This is because each term in
(\ref{eqn:max}) satisfies stochastic dominance for $i=1,\ldots,
b-1$,
\[ X_k\leqst X_m,  T_k\geqst   T_{m}\Rightarrow  T_k(i) - \sum_{j =i}^b X_k(b)  \geqst
   T_{m}(i) - \sum_{j =i}^b X_k(b) .\] Indeed the above terms are
correlated, but they have the same dependency relationship for both
queues $Q_k$ and $Q_m$. Technically, this means that they share the
same copula. The copula $C$ for a multivariate variable $\bfZ$ is
the mapping on the distribution functions such that
\beq\label{eqn:copula}F_{ \bfZ }(\bfz) = C[F_{Z(1)}(z(1)),
F_{Z(2)}(z(2))\ldots].\eeq  By \cite[Thm.
6.B.14]{Shaked&Shanthikumar:book}, under the same copula, we have
the multivariate stochastic order
\[[T_k(b+1), T_k(b)-X_k(b),\ldots] \geqst  [T_{m}(b+1),
T_{m}(b)-X_k(b),\ldots].
\] Hence, their maxima also satisfy stochastic order and  (\ref{eqn:condbatch}) is
true. Since $p_k(x)$ and $p_m(x)$ are decreasing in $x$,
  (\ref{eqn:lemmabatchstuncond}) holds. \qed

\section{Proof of Lemma
\ref{lemma:convex_order_batch}}\label{proof:convex_order_batch} Let
$T'\defeq \lambda T$ be the normalized service time and let $X'(i)$
be i.i.d. Poisson arrivals with unit rate. $\Pbb[B>b|\bfX'=\bfx] $
is given by
\begin{align}\nn &= \Pbb[\max(T'(1), T'(2) + x(1), \ldots, T'(b)+
\sum_{i=1}^{b-1} x(i)) > X'(b)]\\
 &= 1-\Ebb[e^{-\max(T'(1), T'(2) + x(1), \ldots, T'(b)+
\sum_{i=1}^{b-1}
x(i))}].\label{eqn:inproof_convex_order_batch}\end{align} Now, from
convex order,
\begin{align}\nn & T'_k \lcx T'_{m} \Rightarrow \max(T'_k(1), T'_k(2) +
x(1), \ldots, T'_k(b)+ \sum_{i=1}^{b-1} x(i)))\\ &\lcx
\max(T'_{m}(1), T'_{m}(2) + x(1), \ldots, T'_{m}(b)+
\sum_{i=1}^{b-1} x(i))).\end{align} Since
\eqref{eqn:inproof_convex_order_batch} is convex in the argument, it
follows the same order of the service distributions. Since the
convex order is closed under mixtures \cite[Thm.
3.A.12]{Shaked&Shanthikumar:book}, marginalizing over the arrival
times $\bfX'$ preserves the order. Hence, \[ T'_k \lcx T'_{m}
\Rightarrow \Pbb[B_k>b] \leq \Pbb[B_{m}>b],\] which in turn is
equivalent to a stochastic order. \qed

\section{Proof of Theorem \ref{thm:fifo}}\label{proof:fifo}

Given the busy-period size $B=b$, denote the vector of spreads  as
$\bfspread_k $, where the $i^{\tha}$ element is given by
\beq\label{eqn:spread_vector}\spread_k(i)
\defeq  T_k(i)-T_k(i+1) , \quad 1\leq i \leq b-1 .\eeq Note that the elements in the
spread vector  $\bfspread_k$  have identical distributions but are
dependent on one another, unlike the service times of the
infinite-server queue which are independent. We have \[ P_b^\fifo =
\Pbb[\bigcap_{i=1}^{b-1}\{ T_k(i)< X_k(i) + T_k(i+1)\}].\] since
$i^{\tha}$ transaction needs to depart sooner than the
$(i+1)^{\tha}$ transaction.  From the definition of spread vector in
(\ref{eqn:spread_vector}), this is equal to
\begin{align} P_b^\fifo   &=\Pbb[  \bigcap_{i=1}^{b-1}\{
\spread_k(i)< X_k(i) \}], \nn
\\&=\Pbb[T_k(1)<T_k(2)<\ldots]+ \Pbb[  \bigcap_{i=1}^{b-1}\{
0<\spread_k(i)< X_k(i) \}]\nn\\ &= \frac{1}{b!}+ \frac{1}{2} \Pbb[
\bigcap_{i=1}^{b-1}\{ |\spread_k(i)|< X_k(i) \}],\nn \end{align}
since $\spread_k$ is symmetric around zero.  We individually have \[
|\spread_k(i)| \geqst   |\spread_{m}(i)|,\, X_k(i) \leqst
X_m(i),\quad 1\leq i\leq b,\] which implies \[|\spread_k(i)|-X_k(i)
\geqst   |\spread_{m}(i)|-X_m(i),\quad 1\leq i\leq b.\] Since the
spreads $\spread_k(1), \spread_k(2),\ldots$ are correlated,   we use
\cite[Thm. 6.B.14]{Shaked&Shanthikumar:book} to prove  the
multi-variate stochastic order   \beq  |\bfspread_k|-\bfX_k  \geqst
|\bfspread_{m}|-\bfX_m \label{eqn:multorder} ,\eeq  since
  $ |\bfspread_k |-\bfX_k $ and   $|\bfspread_{m}|-\bfX_m$
  share the same copula, defined in (\ref{eqn:copula}).
From \eqref{eqn:multorder},  \[ P_b^\fifo(k) \leqst  P_b^\fifo(m)
\] This implies the order of tracking accuracies  in  (\ref{eqn:fifocond})
by marginalizing over the busy-period sizes since $P_b^\fifo$
decreases in busy period $b$ and the busy periods satisfy stochastic
order, from Lemma \ref{lemma:busyperiod}. \qed

\section{Proof of Theorem \ref{thm:rand}}\label{proof:rand} In order for \eqref{eqn:cond_rand}
to hold, it suffices to show that     \beq  \perm(\bfA_k)|\{B_k=b
\}  \geqst  \perm(\bfA_{m})|\{B_{m}=b \},\label{eqn:permst} \eeq
since the tracking accuracy under random matching is given by
\eqref{eqn:randexpectseries}, and taking expectation over $B_m$ and
$B_k$ preserves the order since   $B_k \geqst B_m$ from
Lemma~\ref{lemma:busyperiod}.  Since the $\perm(\bfA)$ is the number
of matchings for biadjacency matrix $\bfA$, more edges in $\bfA$
implies higher $\perm(\bfA)$. Let $[\alpha_k,\beta_k]$ be the
support of $T_k$ and $[\alpha_m,\beta_m]$ of $T_{m}$. From
(\ref{eqn:amatrix_support})  for $k$, the departure of $i^{\tha}$
arrival has an edge with $j^{\tha}$ arrival, for $1\leq i<j\leq b$
iff.\beq\label{eqn:supportbounds} \alpha_k \leq T_k(i)-
\sum_{a=i}^{j-1} X_k(a)\leq \beta_k.\eeq By definition of support
bound, $T_k(i) \leq \beta$ a.s. Hence, the upper bound in
(\ref{eqn:supportbounds}) always holds. Since $X_k(i)\geq 0$, we
have the probability of edge as \beq \Pbb[A(i,j) =1] =
\bar{F}_{T_k}[\alpha_k+ \sum_{a=i}^{j-1} X_k(a)].\eeq Conditioning
on the same arrival realizations $\bfX_k, \bfX_m = \bfx$,  from the
definition of stochastic dominance,
\begin{align}\nn \bar{F}_{T_k}[\alpha_k+ \sum_{a=i}^{j-1} x(a)]
&\geq \bar{F}_{T_{m}}[\alpha_k+ \sum_{a=i}^{j-1} x(a)]\\ \nn &\geq
\bar{F}_{T_{m}}[\alpha_m+ \sum_{a=i}^{j-1} x(a)],\end{align} when
$\alpha_k\leq \alpha_m$. Now since the functions are decreasing in
$x$ and $X_k \leqst X_m$, the order is preserved on removing the
conditioning. Hence,  (\ref{eqn:permst}) holds implying
\eqref{eqn:cond_rand}.  \qed

\section{Proof of Corollary \ref{cor:rescale}}\label{proof:rescale}
Let $T'_i\defeq \lambda T_i$ for $i=1,2$ be the normalized service
times and let $X'(i)$ be i.i.d.  arrivals with unit rate.  For any
positive  variable  $T'_1$ and $T'_2=c T'_1$ with $0<c <1$, we have
$T'_1\geqst   T'_2$. First consider  FIFO matching rule,  \beq
|\spread_1| \geqst   |\spread_2|= c|\spread_1|,\quad \forall 0<c\leq
1, \eeq and hence, conditions in  Theorem \ref{thm:fifo} for the
order of accuracies under FIFO matching is satisfied.

For random matching rule, let $[\alpha,\beta]$ be the support of
$T'_1$. We have $\alpha > c \alpha$, and hence, the condition in
Theorem \ref{thm:rand} is in fact, violated. We revisit  the
probability of having an edge in the biadjacency matrix $\bfA$
\[ \Pbb_{T'_1}[A(i,j) =1] = \bar{F}_{T'_1}[\alpha+ \sum_{k=i}^{j-1} x(i)].\]
For the service time $T'_2=c T'_1$ with $c <1$, we have
\[ \Pbb_{T'_2}[A(i,j) =1] = \bar{F}_{T'_1}[\alpha+
\frac{1}{c}\sum_{k=i}^{j-1} x(i)]\leq \Pbb_{T'_1}[A(i,j) =1],\] and
hence, the result holds.\qed

\section{Proof of Theorem
\ref{thm:convex_order_opt}}\label{proof:convex_order_opt}

Let $T'\defeq \lambda T$ be the normalized service time and let
$X'(i)$ be i.i.d. Poisson arrivals with unit rate. Given the
busy-period size $B=b$, we have \begin{align}\nn  P_b^\fifo| \{B=b\}
&=  \Pbb[\bigcap_{i=1}^{b-1}\{ T'(i)- T'(i+1)< X'(i) \}]\\ &=
\Ebb[\exp{(-\sum_{i=1}^b(T'(i)-T'(i+1))^+)}]\nn
\\&=\Ebb[\exp{(-\sum_{i=1}^b a_{i,\pi} T'(i))}| \Pibf(\bfT') =\pibf ] \nn
,\end{align} where $a_{i,\pi} = 0, \pm 1$ are fixed coefficients
conditioned on the event that the service times $\bfT'$ follow a
certain permutation $\pibf$.  Now $\exp{(-\sum_{i=1}^b a_{i,\pi}
T'(i))}$ is a concave function of $\sum_{i=1}^b a_{i,\pi} T'(i)$ and
all permutations $\pibf$ of the service times are equiprobable at
both the queues (since all the service times are i.i.d.).

 On the lines of \cite[Thm.
3.A.19]{Shaked&Shanthikumar:book}, we can show that when $\bfT'_k$
and $\bfT'_m$ are conditioned on the same permutation $\pibf$,
\[ T'_k\lcx T'_{m} \Rightarrow \sum_{i=1}^b a_{i,\pi} T'_k(i) \lcx
\sum_{i=1}^b a_{i,\pi} T'_m(i).\] Hence, \[ T'_k\lcx T'_{m}
\Rightarrow P_b^\fifo(k)| \{B_k=b\}  \geq P_b^\fifo(m)| \{B_m=b\}.\]
Since $P_b^\fifo$ is decreasing  in $b$ and the busy-period sizes at
$k$ and $m$ follow the stochastic order, the order carries through
when we marginalize over the busy-period sizes. \qed

\section{Proof of Theorem~\ref{thm:ps}}\label{proof:ps}

Let $\bfT_k$ and $\bfT_{m}$ be the sojourn times of the jobs in the
two queues.  The sojourn times satisfy
 \[ X_k \leqst X_m, \,
 J_k \geqst J_m \Rightarrow T_k(i) \geqst  T_{m}(i).\]
 Now $\bfT_k$ and $\bfT_{m}$  are correlated,
unlike the infinite-server case. However, $\bfT_k$ and $\bfT_{m}$
have the same  copula since they are both processor sharing queues
 and by \cite[Thm. 6.B.14]{Shaked&Shanthikumar:book}, \[ X_k \leqst
X_m, \,
 J_k \geqst J_m \Rightarrow \bfT_k  \geqst  \bfT_{m}.\] On lines of
 Lemma~\ref{lemma:busyperiod}, \[X_k \leqst
X_m, \,\bfT_k  \geqst  \bfT_{m} \Rightarrow B_k \geqst B_m.\]
  Note that the lower bound of support of each sojourn
time is the same as the job lengths.   On lines of~\ref{proof:rand},
 \eqref{eqn:ps_rand} holds. \qed

\section{Proof of Theorem~\ref{thm:inf_ps}}\label{proof:inf_ps}

We first provide a result that under the stochastic dominance
assumption, the sojourn times of the processor-sharing queue
dominate those of the infinite-server queue.

 \begin{proposition}\label{prop:multorder_inf_ps}
 \textsc{(Sojourn Times in Infinite Server and Processor Sharing
Queues).} We have \beq  J_\psq \geqst T_\inft \Rightarrow \bfT_\psq
\geqst \bfT_\inft .\label{eqn:multorder_inf_ps}\eeq
\end{proposition}

\bprf The multivariate ordering is implied by the conditional
ordering
\[T_\psq(i) |\bigcap_{k=1}^{i-1} \{ T_\psq(k)=t_k\}\geqst T_\inft.\] Now the sojourn times
$T_\psq(i)$ at the processor-sharing queue are at least the job
lengths with probability 1. Hence,  upon any conditioning
\[T_\psq(i) |\bigcap_{k=1}^{i-1}\{ T_\psq(k)
=t_k\} \geqst J_\psq(i), \quad i=1,2,\ldots.\] Hence, the result in
\eqref{eqn:multorder_inf_ps} holds.\eprf

The above result follows the intuition that when larger jobs are
arriving to the processor-sharing queue than to the infinite-server
queue, the sojourn times in the processor-sharing queue are longer.
However, the converse is not always true since  even longer jobs can
have shorter sojourn times in the infinite-server queue due to
simultaneous processing of the jobs.

We now use the above proposition to provide a result on the
busy-period sizes. From \eqref{eqn:multorder_inf_ps}, we have the
multivariate stochastic order. Now,  \[ \bfX_\psq \leqst \bfX_\inft,
\bfT_\psq  \geqst \bfT_\inft \Rightarrow B_\psq \geqst B_\inft,\] on
lines of Lemma~\ref{lemma:busyperiod}. On lines of
Theorem~\ref{thm:rand}, we have \eqref{eqn:inf_ps}. \qed

\end{appendix}

%
%
%

\end{document}